\documentclass[11pt, A4, leqno]{article}
\usepackage{graphicx}
\usepackage{latexsym}
\usepackage{amsfonts}
\usepackage{amsmath}
\usepackage{xcolor}
\numberwithin{equation}{section}

\newcommand{\qed}{\hfill $\Box$}

\newtheorem{theorem}{Theorem}[section]

\newtheorem{example}[theorem]{Example}

\newtheorem{remark}[theorem]{Remark}

\begin{document}

\title{ Turnpike Property and Convergence Rate for an Investment and Consumption
Model}
\author{Baojun Bian\thanks{Department of Mathematics,
         Tongji University, Shanghai 200092, China.
bianbj@tongji.edu.cn, Research of this author was supported by
NSFC 11371280.}, Harry Zheng\thanks{ Department of
Mathematics, Imperial College, London SW7 2BZ, UK. 
h.zheng@imperial.ac.uk. Tel: +44 2075948539}}
\date{}
\maketitle

\noindent{\bf Abstract.}
 We discuss the turnpike property for optimal investment and consumption problems. We find there exists a threshold value that determines the turnpike property for investment policy. The threshold value only depends on the Sharpe ratio, the riskless interest rate and the discount rate. We show that
if utilities  behave asymptotically like power utilities  and satisfy some simple relations with the threshold value,  then  the turnpike property for investment holds. There is in general no turnpike property for consumption policy. We also provide the rate of convergence  and illustrate the main results with examples of  power and non-HARA utilities and numerical tests.




\section{Introduction}\label{intro}

It is well known  that the optimal portfolio strategy for terminal wealth (amount of money) utility maximization  problems in a Black-Scholes market can be approximated by a wealth-  and time-independent strategy  if the planning horizon is distant  (investment over the long run) and the terminal wealth utility  behaves asymptotically like a power utility.  This is called the {\it turnpike property} for investment,
see Back et al. (1999),  Bian and Zheng (2015),  Cox and Huang (1992), and Huang and Zariphopoulou (1999) for expositions on the topic. It is highly interesting to know if the turnpike property still holds for optimal investment and consumption problems. Consider the following  utility maximization problem:
\begin{equation} \label{primal_problem}
\sup_{\pi, c} E\left[\int_0^T e^{-\delta t} U_2(c_t)dt +e^{-\delta T}U_1(X_T)\right],
\end{equation}
 where $\delta$ is a discount factor,
$T$ is the planning horizon,  $X_T$ is  the wealth at time $T$  and $X$ is a wealth process satisfying 
\begin{equation} \label{wealth}
dX_t = rX_tdt + X_t\pi_t \sigma (\theta dt + dW_t) - c_tdt, \; t\geq 0,
\end{equation}
with the initial wealth $X_0=x_0$,  $r$ is the riskless interest rate,
 $\theta=\sigma^{-1}(\mu-r)$ is the Sharpe ratio,  $\mu$  and $\sigma$ are the growth 
  and volatility rates of a risky asset, $W$ is a standard Brownian motion,
$\pi$ and $c$ are  portfolio and consumption processes, $U_i$, $i=1,2$, are utilities for wealth and consumption. Assume $A(x,t)$ is the optimal amount of money invested  in the risky asset at time $t$ with wealth $x$. We say the problem (\ref{primal_problem}) has the turnpike property for investment if $A(x,t)$ is approximately a linear function of $x$ when $T$ is distant.
If $U_i$ are the same power utility function $(1/p)x^p$, then
$$ A(x,t)={\theta\over \sigma (1-p)}x.$$
Jin (1998) proves that if utilities $U_i$ behave asymptotically like  power utilities $(1/p_i)x^{p_i}$ for large wealth $x$, then
 the optimal portfolio $\pi$  and  the optimal consumption $c$ at any fixed time $t$ are close to those derived with  power utilities $(1/ p_i)x^{p_i}$   in the  absolute or mean squared norm  if the investment horizon $T$ is distant.
Back et al. (1999) discuss portfolio turnpikes for optimal terminal wealth (called consumption in their paper)  problems and  show that the turnpike property does not hold in the presence of consumption (called consumption withdrawal in their paper) with a counter-example using a shifted power utility.
Little is known in the literature if $A(x,t)$ is still approximately a linear function of $x$  for general utilities $U_i$ when $T$ is distant,  even less so for the limiting behaviour of the optimal consumption $c(x,t)$.

The objective of this paper is to identify the conditions for general utilities $U_i$ under which the optimal portfolio strategy can  be approximated by a wealth-  and time-independent strategy  if the planning horizon is distant. When utilities $U_i$ behave asymptotically like power utilities $(1/p_i)x^{p_i}$ at large wealth $x$, we can give an affirmative answer on whether or not the turnpike property holds.
The main contributions of the paper are  that we find a threshold value  that characterizes precisely the conditions for utilities to have the turnpike property, that we show there is no turnpike property in general for the optimal consumption, and that we estimate  the convergence rate of the optimal   trading strategies to the limiting ones,   which, to the best of our knowledge, is absent in the literature,  see Theorems \ref{Main-1}, \ref{Main-1a} and \ref{Main-2}.

We next highlight the main results of the papers. In the literature the conditions are normally imposed on utilities $U_i$ for the turnpike property. For example, assume $U_i$
 are continuously differentiable, increasing and strictly concave, satisfying $\lim_{x\to\infty}U_i'(x)=0$ and
\begin{equation} \label{V-1}
 \lim_{x\to \infty} {U_i'(x) \over  x^{p_i-1}} = 1,
\end{equation}
for $i=1,2$, where $p_i<1$  and $U_i'$ are derivatives of $U_i$.
 Condition (\ref{V-1})  means that utilities $U_i$ behave like power utilities $(1/p_i)x^{p_i}$ if the wealth level $x$ is large and is equivalent to
\begin{equation}  \label{eqn10}
\lim_{y\to0} {V_i'(y)\over y^{q_i-1}} = -1,
\end{equation}
where  $q_i:=p_i/(p_i-1)<1$ and $V_i$ are the  dual functions of $U_i$,  defined by
\begin{equation}\label{dual_func_V_i}
V_i(y):=\sup_{x\geq 0}\{U_i(x)-xy\}
\end{equation}
for $y\geq 0$.\footnote{A utility function $U$ and its dual function $V$  are equivalent and can be recovered by each other from the relations $V(y)=\sup_{x>0}(U(x)-xy)$ and $U(x)=\inf_{y>0}(V(y)+xy)$.  For example, if $U$ is a power utility $U(x)=(1/p)x^p$ for $x>0$ and $p<1$, then its equivalent dual function is $V(y)=-(1/q)y^q$ for $y>0$ and $q=p/(p-1)$.
Furthermore, $U_i(x)=V_i(y)+xy$ if and only if  $U'_i(x)-y=0$. Therefore, from
$$ {U_i'(x)\over x^{p_i-1}}=
 \left({y^{q_i-1}\over -V_i'(y)}\right)^{p_i-1}
$$
and $U_i'(\infty)=0$, we get the equivalence of
conditions (\ref{V-1}) and (\ref{eqn10}).
}

 The key benefit of using dual utilities  $V_i$, instead of utilities $U_i$, is the following:  When the stochastic control method is used  to  solve  investment and consumption problems, the optimal strategies can be expressed by  some functions of derivatives of  a solution to  a  nonlinear  partial differential equation (PDE) which is difficult to solve and analyse.
 Thanks to the dual stochastic  control method, we demonstrate that  the optimal strategies can be characterized by some functions of  derivatives of a solution to a linear PDE and  have representations in terms of dual utilities $V_i$, which makes it feasible to derive the turnpike property and estimate the convergence rate.

 Theorem \ref{Main-1} states that if
  (\ref{eqn10}) holds\footnote{The result of Theorem \ref{Main-1} can be stated  equivalently in terms of $p_i$ defined in (\ref{V-1}), that is, if $p_1>p^*$ or $p_2\geq p^*$, where  $p^*=q^*/(q^*-1)\geq 0$, then the turnpike property holds
and the optimal amount of investment can be approximated by
$ \lim_{T\rightarrow \infty}A(x,t)=(\theta/\sigma)(1-\max\{p_1,p_2\})^{-1}x$.}
 with  $q_1<q^*$ or $q_2\leq q^*$, where $q^* <  0$ is a threshold value,\footnote{Note  that $q^*$ only depends on the market price of risk $\theta$, the riskless interest rate $r$ and the utility discount rate $\delta$. }
 given by
\begin{equation}
q^*=\alpha - \sqrt{\alpha^2 + {2\delta \over \theta^2}}, \label{threshold_value}
\end{equation}
where $\alpha={1\over 2}+ {r-\delta \over \theta^2}$,
 then the turnpike property for investment holds, that is,
\begin{equation}\label{turnpike}
\lim_{T\rightarrow \infty}A(x,t)=\frac{\theta}{\sigma}(1-\min\{q_1,q_2\})x,
\end{equation}
 which means the optimal amount of investment $A(x,t)$ can be approximated by  $\frac{\theta}{\sigma}(1-\min\{q_1,q_2\})x$ when the investment horizon $T$ is distant.

  Theorem \ref{Main-1a} states that if dual marginal utilities $V_i'$ converge to dual marginal power utilities $-y^{q_i-1}$ at certain speed and $q_1<q^*$ or $q_2<q^*$, then
the speed of convergence of the optimal investment  strategy to its  limiting strategy in (\ref{turnpike}) is exponentially fast.

 Theorem \ref{Main-2} states that if   (\ref{eqn10}) holds with $q_1\geq q^*$ and $q_2> q^*$,  then $A(x,t)$ converges to a nonlinear function of $x$ when $T\to\infty$ for general utilities, in other words, the turnpike property does not hold in the classical sense.  However, there is a notable exception if the  consumption utility $U_2$ is a power utility $(1/p_2)x^{p_2}$ with $0<p_2<1$, in that case $A(x,t)$ still converges to a linear function of $x$, given by $(\theta/\sigma)(1-q_2)x$.

 We now illustrate the results of Theorems \ref{Main-1} to \ref{Main-2} with both utilities being    power utilities.
  It is well known that the optimal amount of investment $A(x,t)$   is a linear function of the wealth $x$ if there is only terminal wealth utility $U_1$ ($U_2=0$) or only consumption utility $U_2$ ($U_1=0$) or the same utilities $U_1=U_2$. For   different power utilities $U_1$ and $U_2$,  $A(x,t)$ is a nonlinear function of $x$. It is not clear  how  behaves if the investment horizon $T$ is distant. Thanks to Theorems \ref{Main-1} to \ref{Main-2}, we conclude that  the turnpike property for investment  essentially holds, that is,
 \begin{equation}\label{A(x,t)}
   \lim_{T\to\infty} A(x,t) = \left\{\begin{array}{ll}
 \frac{\theta}{\sigma}(1-\min\{q_1,q_2\})x& \mbox{if  $q_1<q^*$ or $q_2\leq q^*$}\\
 {\theta\over \sigma}(1-q_2)x & \mbox{if $q_1>q^*$ and $q_2>q^*$},
 \end{array}\right.
 \end{equation}
 where $q_i=p_i/(p_i-1)<0$. Furthermore, the convergence speed is exponentially fast if $q_1<q^*$ or $q_2<q^*$. If $q_1=q^*$ and $q^*<q_2<0$ then $A(x,t)$ converges to a nonlinear function of $x$, see Theorem \ref{Example-1} for details. This is a new result even for power utilities, not to mention  our main theorems cover general utilities.

We next give a numerical test. The data used are the Sharpe ratio  $\theta=0.2$, the discount rate $\delta=0.02+(1/2)r$, which gives the threshold value $q^*=-1$,  the volatility rate $\sigma=0.2$,  the riskless interest rate $r=0.02$, 0.06, 0.1, which gives the discount rate $\delta=0.03$, 0.05, 0.07,  respectively,  the time horizon $T-t=1$, 2, 5, 10, 25, 50, and 100 years.  We discuss three cases: 1) $q_1=-1/2$ and $q_2=-2$; 2) $q_1=-2$ and $q_2=-1/2$; 3) $q_1=-1/2$ and $q_2=-1/4$. From (\ref{A(x,t)}), we know the optimal proportion of wealth $\pi^*(x,t):=A(x,t)/x$ converges to the Merton portfolio $\pi_M(x,t)=3$ in cases 1 and 2 and converges to $\pi_M(x,t)=1.25$ in case 3  as $T-t$ tends to $\infty$.

Table \ref{power_table} lists the values of  optimal portfolio $\pi^*(x,t)$ for different  time horizons $T-t$ and riskless interest rates $r$. It is clear that as $T-t$ tends to infinite, the exact optimal portfolio converges to the Merton portfolio in (\ref{A(x,t)}). However, the speed of convergence is not consistent for different power utilities: it is  fast in  cases 1 and 3, but  slow in case 2 (when $T-t=200$ the exact optimal portfolio values are 2.9789 for $r=0.02$, 2.9981 for $r=0.06$ and 2.9998 for $r=0.10$). Back et al. (1999) claim with numerical examples  that one has the turnpike property only when the investment horizon is very long in a low interest rate economic environment.  Case 2 is in line with that finding, but cases 1 and 3  show that the convergence is still reasonably fast.    In fact, we can find the exponential convergence rates for all cases. For example, in case 1, convergence rate is 0.055 for $r=0.02$, 0.075 for $r=0.06$ and 0.095 for $r=0.10$. The higher the interest rate, the faster the convergence of optimal portfolio to its limiting portfolio.

\begin{table}
\begin{center}
\begin{tabular}{|l|c|ccccccc|}
\hline
$q_1, q_2, \pi_M$ &$r\backslash T-t$ & 1&2&5&10&25&50&100\\ \hline
 $q_1=-1/2$& 0.02&2.6075&	2.7195&	2.8345&	2.9014&	2.9655&	2.9919&	2.9995\\
$q_2=-2$  &0.06&2.6125&	2.7271&	2.8461&	2.9152&	2.9771&	2.9966&2.9999\\
 $\pi_M=3$ &0.10& 	2.6174&	2.7344&	2.8569&	2.9273	&2.985	&2.9986&3.000\\	
\hline
$q_1=-2$  &0.02 &2.6099&	2.348&	1.9557&	1.7538	&1.6742&	1.8119&	2.4955\\
$q_2=-1/2$  &0.06&2.6149&	2.362	&1.9924	&1.8107	&1.7963&	2.1486	&2.8421\\
$\pi_M=3$ &0.10&2.6198&	2.3759&	2.0297	&1.8739&	1.9555&2.4723&	2.9532\\
\hline
$q_1=-1/2$ &0.02 &1.3928&	1.3458	&1.2944&	1.2714&	1.2571	&1.2527&	1.2507\\
$q_2=-1/4$  &0.06&1.3927&	1.3456	&1.2942	&1.2712&	1.2567&	1.2522	&1.2504\\
$\pi_M=1.25$  &0.10&1.3927&1.3455&	1.294&	1.2708&	1.2563	&1.2518&	1.2502\\
\hline
\end{tabular}
\end{center}
\caption{Optimal portfolios $\pi^*(x,t)$ with different    time horizons $T-t$,
interest rates $r$, and dual power utilities $V_i(y)=-(1/q_i)y^{q_i}$,
or equivalently,
power utilities $U_i(x)=(1/p_i)x^{p_i}$, where
$p_i=q_i/(q_i-1)$ for $i=1,2$. For $q=-1/4,-1/2,-2$, the corresponding $p=1/5, 1/3, 2/3$, respectively. The threshold value $q^*=-1$, the corresponding $p^*=1/2$. The other data used are the Sharpe ratio
 $\theta=0.2$, the discount rate $\delta=0.02+(1/2)r$ and the volatility rate $\sigma=0.2$. $\pi_M$ is the Merton portfolio.}
\label{power_table}
\end{table}


   The optimal consumption strategies are in general nonlinear functions of $x$ for any $q_1$ and $q_2$, see Theorems \ref{Main-1} and  \ref{Main-2}.
    The economic reason  of non-existence of the turnpike property of consumption  is that the utility  from consumption is over the whole investment period and the initial  wealth and consumption  may be small, which implies even though the consumption utility behaves like a power utility with large consumption, but  at the beginning  one has to use the specific consumption utility in deciding the optimal consumption strategy that may  be a nonlinear function of the wealth.
  This phenomenon is present only when there are both terminal and consumption utilities.

 Theorem \ref{Main-3} states that if
\begin{equation}\label{V-wC0}
\lim_{y\to0}R_i(y):=\left(-{yV_i''(y)\over V_i'(y)}\right) = 1-q_i,\quad i=1,2,
\end{equation}
and $q_1, q_2 <q^*$ and some other conditions, then  the turnpike property (\ref{turnpike}) holds.\footnote{
Condition (\ref{V-wC0}) is equivalent to the limits of the Arrow-Pratt coefficients of relative risk aversion of utilities $U_i$ being $1-p_i$, that is,
\begin{equation}\label{V-2}
\lim_{x\to \infty}\left(-{xU_i''(x)\over U_i'(x)}\right)   = 1-p_i,\quad i=1,2,
\end{equation}
and $p_1,p_2>p^*$. This is due to the dual relation of $U_i$ and $V_i$ defined in (\ref{dual_func_V_i}), which implies
that if $y=U_i'(x)$ then $x=-V_i'(y)$ and $U_1''(x)=-1/V_1''(y)$,  and therefore the equivalence of conditions (\ref{V-wC0}) and (\ref{V-2}).
}
Conditions (\ref{eqn10}) and (\ref{V-wC0})  are  in general not implied by each other, see Footnote \ref{rk4.11} for examples. Condition (\ref{V-wC0}) may be relaxed further for the turnpike property, in particular, $V_1$ is only required to be regularly varying at zero, see Theorem \ref{Main-4} for details.

Finally we want to emphasize that the 
methodology presented in this paper depends on Black-Scholes market with constant investment opportunities.  For turnpike problems with only terminal consumption, models with stochastic investment opportunities (incomplete markets) have been studied in Guasoni et al. (2014) and  Robertson and Xing (2017), see detailed discussions and other references in these two papers.

The rest of the paper is organized as follows. In Section 2 we formulate the model, classify different cases for  the turnpike property   when both $U_1$ and $U_2$ are power utilities (Theorem \ref{Example-1}), state and discuss the main results of the paper (Theorems \ref{Main-1}--\ref{Main-4}).
In Section 3 we apply the main theorems to two examples with power and non-HARA utilities and perform some numerical tests and analysis. Section 4 concludes.  The appendix discusses the primal and dual approaches to solving utility maximization problems, derives the optimal investment and consumption strategies, and gives the detailed  proofs of all theorems.


\section{Turnpike property and convergence rate}

Consider a financial market consisting of one riskless  asset and one risky asset.
The  price process of  $S$ of the  risky asset is modelled by
$$
dS_t=S_t(\mu dt+ \sigma dW_t)
$$
for $0\leq t\leq T$,
where $\mu$ is the return rate and $\sigma$ the volatility rate of the risky asset, both are positive constants, and  $W$ is  a standard Brownian motion
 on a complete probability space
$(\Omega,\mathcal{F},P)$, endowed with a natural filtration
$\{\mathcal{F}_t\}$ generated by $W$.
The  wealth process $X$  satisfies the SDE  (\ref{wealth}), that is,
 \begin{equation*} 
dX_t = rX_tdt + X_t\pi_t \sigma (\theta dt + dW_t) - c_tdt,
\; X_0=x_0,
\end{equation*}
where $r>0$ is the riskless interest rate, $\theta=\sigma^{-1}(\mu-r)$ is the Sharpe ratio,   $\pi$ is a proportional portfolio process and $c$ a nonnegative consumption rate process, satisfying the standard measurability and integrability conditions.
The wealth process $X$ is driven by only one stock.
 The extension to multiple correlated
  stocks is straightforward.
 We therefore focus on the  model (\ref{wealth}).

Consider the  utility maximization problem (\ref{primal_problem}).
Assume $U_i$, $i=1,2$, are continuously differentiable, strictly increasing, and strictly concave functions on  $R_+:=(0,\infty)$,  satisfy $U_i(0)=0$,  $U_i'(0)=\infty$ and $U_i'(\infty)=0$, and
$ U_i(x)\leq  C(1+x^p)$ for $x\geq 0$ and some constants  $C>0$ and $0<p<1$.
Note that the assumption $U_i(0)=0$ is not needed in Theorem \ref{Example-1} but is required for all other results.

To simplify the notation, we define $\tau=T-t$, the time horizon. Then $T\to \infty$ is equivalent to $\tau\to\infty$. We still use $t$ to represent a time horizon variable, instead of $\tau$.

Using the dual stochastic control method, we can show that the optimal amount of investment and consumption rate are given by
\begin{equation} \label{ICS-1}
A(x,t)=x\pi^*(x,t)={\theta\over \sigma}\left(J_1(y,t)+\int_{0}^tJ_2(y,\tau)d\tau  \right) ,\ \ \  C(x,t)= c^*(x,t)=-V'_2 (y),
\end{equation}
where $y  =y(x,t)$ is the solution to the budget constraint equation
\begin{equation}\label{bce}
x= I_1(y,t)+\int_{0}^t I_2(y,\tau)d\tau.
\end{equation}
and,  for i=1, 2,
\begin{eqnarray*}
I_i(y,t)&=&\frac{e^{\beta t}}{2\sqrt{\pi}}\int_{-\infty}^\infty
e^{-\frac{\eta^2}{4}-(\alpha-1) a\sqrt{t}\eta}|V_i'(ye^{a\sqrt{t}\eta})|d\eta\\
J_i(y,t)&=&\frac{e^{\beta t}}{2\sqrt{\pi }}\int_{-\infty}^\infty
e^{-\frac{\eta^2}{4}-(\alpha-1) a\sqrt{t}\eta}ye^{a\sqrt{t}\eta} V_i''(ye^{a\sqrt{t}\eta})d\eta,
\end{eqnarray*}
and $\alpha=\frac{1}{2}+\frac{r-\delta}{\theta^2}$, $a=\frac{\theta}{\sqrt{2}}$,
$\beta=-a^2 \alpha^2-\delta $.
Define
$$
\lambda=\lambda(q):
=\beta+(\alpha-q)^2a^2=\frac{1}{2}\theta^2 q(q-1)-rq + \delta(q-1)
$$
 for  $q< 1$.  Noting $\lambda(0)=-\delta  < 0$  and $\lambda(1)=-r < 0$, we conclude that there is a unique root $q^* <0$ to the equation $\lambda(q)=0$ for $q< 1$,  given by
(\ref{threshold_value}),
and
 $\lambda(q)<0$ for $q^*<q<1$ and $\lambda(q)>0$ for $q<q^*$.

 \begin{remark}
 In this paper we assume there is only one risky asset in the market. It is straightforward to extend the results to a complete market with $n$ assets and $n$ standard Brownian motions.  In that case, market price of risk $\theta$ becomes a vector: $\theta=\sigma^{-1}(\mu-r{\bf 1})$, where $\sigma$ is a stock  volatility matrix, $\mu$ is a stock growth rate vector,  and ${\bf 1}$ is a vector with all components 1. The only change we need to do in the definition of $\lambda(q)$ is to replace $\theta^2$ with $\|\theta\|^2=\theta^T\theta$. The optimal amount of money invested in asset $i$ is given by
 $A_i(x,t)=x{\bf e_i}^T \pi^*(x,t) = {\bf e_i}^T (\sigma^T)^{-1}\theta \left(J_1(y,t)+\int_{0}^tJ_2(y,\tau)d\tau  \right)$, where ${\bf e_i}$ is a vector with all components 0 except the $i$th component which is 1. All results for one risky asset model still hold for this multiple risky asset model. For this reason, we only discuss one risky asset in this paper.
 \end{remark}
 
 \begin{remark}
We apply the dual stochastic control method to derive (\ref{ICS-1}) and (\ref{bce}), which are the key relations for the turnpike property. Since the dynamic programming equation (the HJB equation) is used for solving the primal and dual problems, the model must be Markovian and time-consistent, that is, with the current methodology, we cannot cover the model with random coefficients nor mean-variance problems. However, it is possible to extend the model from a complete market unconstrained setting to the one with control constraints (no short selling or no trading of some assets, etc.), that is, $\pi_t\in K$ where $K$ is a closed convex cone in $R^n$.  This is because the dual HJB equation   (see (\ref{dual_HJB})) is still a linear PDE and can be solved with a Feynman-Kac representation. The only change for the dual HJB equation (\ref{dual_HJB}) is to replace $\theta^2$ with $\|\hat\theta\|^2$, where $\hat\theta := \theta + \sigma^{-1}\hat \pi=\sigma^{-1}(\mu - r{\bf 1}+\hat\pi)$ and $\hat\pi$ is the unique minimizer of a quadratic function $f(\tilde\pi):=\|\theta+\sigma^{-1}\tilde \pi\|^2$ over $\tilde\pi\in \tilde K: =\{p: p' v \geq 0,\;\forall v\in K\}$, the positive polar cone of $K$ in $R^n$.  In the presence of control constraints, one cannot use the martingale representation theorem to find the optimal control as in a complete market setting, but may derive them using the stochastic control approach, see further details in Bian et al. (2011) and Bian and Zheng (2015). Since relations (\ref{ICS-1}) and (\ref{bce}) essentially hold for closed convex cone constrained problems, we expect the results for the turnpike property still hold and therefore only focus on the unconstrained case in this paper. 
 \end{remark}

Equation (\ref{bce}) is a budget constraint in which the initial wealth $x$ is used to finance the optimal terminal wealth (the first term) and the optimal total consumption (the second term).
We want to show one term dominates the other, that is, one term tends to 0 and the other tends to $x$ as  $t\to\infty$.  Once this is decided, the limiting properties for $A(x,t)$ and $C(x,t)$ are immediate from (\ref{ICS-1}). It turns out $q^*$ is a threshold value that determines which term dominates in  (\ref{bce}).

 The next theorem  characterizes the turnpike property when both utilities are power utilities.

\begin{theorem}\label{Example-1} Let $U_i(x)=(1/p_i)x^{p_i}$ and $p_i<1$ for $i=1,2$ (if $p_i=0$ then $U_i(x)=\ln x$).
Define $q_i:=p_i/(p_i-1)  <1$ and $\lambda_i:=\lambda(q_i)$, $i=1,2$. We have
\begin{enumerate}
\item If  $q_1<q^*$ or $q_2\leq q^*$, then
\begin{eqnarray*}
\lim_{t\rightarrow \infty}A(x,t)&=&\frac{\theta}{\sigma} (1-\min\{q_1, q_2\})x,\\
\lim_{t\to\infty} R(t) C(x,t)&=&x^{\frac{ q_2-1}{\min \{q_1,q_2\}-1}},
\end{eqnarray*}
where
\begin{equation} \label{R(t)}
R(t)=\left\{\begin{array}{lll}
e^{\lambda_1 t}+\frac{e^{\lambda_2 t }-1}{\lambda_2}, & q_1=q_2, \\
e^{\frac{q_2-1}{q_1-1}\lambda_1 t}, & q_1<q_2, \\
\frac{e^{\lambda_2 t}-1}{\lambda_2}, & q_1>q_2,
\end{array}\right.
\end{equation}
 and $(e^{\lambda_2t}-1)/\lambda_2=t$ if  $\lambda_2=0$.
\item If  $q_1> q^*$ and $q_2> q^*$, then
\begin{eqnarray*}
\lim_{t\rightarrow \infty}A(x,t)&=&\frac{\theta}{\sigma }(1-q_2)x.\\
\lim_{t\rightarrow \infty}C(x,t) &=&-\lambda_2 x,
\end{eqnarray*}
\item If  $q_1= q^*$ and $q_2> q^*$, then
\begin{eqnarray*}
\lim_{t\rightarrow \infty}A(x,t)&=&\frac{\theta}{\sigma }[(1-q_1)Y^{q_1-1}-(1-q_2)\frac{1}{\lambda_2}Y^{q_2-1}],\\
\lim_{t\rightarrow \infty}C(x,t) &=&Y^{q_2-1},
\end{eqnarray*}
where  $Y$ is the unique solution to the equation
$y^{q_1-1}-\frac{1}{\lambda_2}y^{q_2-1}=x$.
\end{enumerate}
\end{theorem}

Theorem \ref{Example-1} can be recovered from Theorems \ref{Main-1} and \ref{Main-2} for general utilities  if $0<p_i<1$ for $i=1,2$, but not if  $p_i\leq 0$ as the condition $U_i(0)=0$ is not satisfied.
However, due to the homothetic property of the power utility,  Theorem~\ref{Example-1} can be proved without using the condition $U_i(0)=0$,  see Appendix for a rigorous proof.
 
We outline the key idea of the proof of Theorem \ref{Example-1}.
From (\ref{ICS-1}) we get
\begin{equation}\label{control-i}
A(x,t)={\theta\over \sigma}\left((1-q_1)e^{\lambda_1 t}y^{q_1-1} + (1-q_2) \frac{e^{\lambda_2 t}-1}{\lambda_2}y^{q_2-1}\right), \quad
C(x,t)=y^{q_2-1},
\end{equation}
where $y$ is the solution to the equation
 \begin{equation}\label{relationship}
x=e^{\lambda_1 t} y^{q_1-1} +\frac{e^{\lambda_2 t }-1}{\lambda_2} y^{q_2-1}.
\end{equation}
The initial wealth $x$ is used to finance the optimal terminal wealth and the optimal total consumption. The relation of $q_1$ and $q_2$ determines which term  in (\ref{relationship}) dominates  as $t$ tends to $\infty$. If the first term dominates, that is, if the first term tends to $x$ and the second term tends to 0 as $t$ tends to $\infty$, then essentially all initial wealth is used for maximizing  the utility of the terminal wealth, and the optimal investment strategy is therefore  determined approximately by the utility of terminal wealth. This can be easily seen from (\ref{relationship}) and (\ref{control-i}). It is more complicated in finding the limiting relation for the optimal consumption.
Similar discussions apply to the cases when the second term dominates or neither term dominates.

We next give some discussions on the results of Theorem \ref{Example-1}.

In case 1, since $q_1<q^*$ or $q_2\leq q^*$, we have $\lambda_1>0$ or $\lambda_2\geq 0$. This leads to $\lim_{t\rightarrow \infty}R(t)=\infty$  and $\lim_{t\rightarrow \infty}C(x,t)=0$  for any $x>0$, which implies the initial consumption $C(x,t)$ should be close to zero and  $R(t)$ is the speed of the consumption tending to zero when the horizon $t$ is distant. From the budget constraint  (\ref{relationship}) and  $\lambda_1>0$ or $\lambda_2\geq 0$, we have the marginal utility $y=u_x(x,t)$ tends to $\infty$ as $t\to\infty$ for any fixed $x$, which implies one should invest  in the risky asset to increase the wealth level and therefore the overall utility.

In case 2, since $q_1>q^*$ and $q_2> q^*$, we have $\lambda_1<0$ and $\lambda_2< 0$, which implies the marginal utility $y=u_x(x,t)$ is bounded away from 0 and $\infty$ as $t\to\infty$ (we can prove it by contradiction argument as follows:  if $y\to 0$ (or $\infty$) as $t\to\infty$, then the  right side of (\ref{relationship}) tends to $+\infty$ (or 0), but $x$ on the left side is a positive number). The second term  in (\ref{relationship}) dominates and the optimal trading strategies are determined approximately  by the utility of consumption only, which is a  Merton strategy.

In case 3,  since $q_1=q^*$ and $q_2>q^*$, we have $\lambda_1=0$ and $\lambda_2< 0$, which again implies the marginal utility $y=u_x(x,t)$ is bounded away from 0 and $\infty$ as $t\to\infty$. Since neither term  in (\ref{relationship}) dominates, the optimal trading strategies  are determined  jointly by
 the utilities of terminal wealth and  consumption, which results in the optimal  investment strategy  being a nonlinear function of wealth.


 We now state and discuss the main results of the paper for general utilities.
\begin{theorem} \label{Main-1}
Let $q_1<q^*$ or  $q_2 \leq q^*$.
Assume that $V_i\in C^1(R_+)$ and satisfy (\ref{eqn10}) for $i=1,2$.
Then for any $x\in R_+$, we have  the turnpike property (\ref{turnpike}).
If, for some $\hat q <0$,
\begin{equation}\label{V-C-C}
\lim_{y\rightarrow \infty} \frac{V'_2(y)}{y^{\hat q-1}}=-1,
\end{equation}
then, we have
\begin{equation}\label{consumption}
\lim_{t\rightarrow \infty}R(t)^{\frac{\hat q-1}{q_2-1}}C(x,t) =x^{\frac{ \hat q-1}{\min \{q_1,q_2\}-1}},
\end{equation}
where $R(t)$ is defined in (\ref{R(t)})
\footnote{
 The limits in conditions  (\ref{eqn10}) and (\ref{V-C-C})  are $-1$ which can be replaced by some constants. Specifically, if
 there exist $k_i>0$, such that $\lim_{y\rightarrow 0}\frac{V_i'(y)}{y^{q_i-1}}=-k_i$,
then for $x\in R_+$, we still have the turnpike property
(\ref{turnpike}).
If, for some $\hat q <1$, $k\geq 0$,
$\lim_{y\rightarrow \infty} \frac{V'_2(y)}{y^{\hat q-1}}=-k$,
then, we have
$\lim_{t\rightarrow \infty}\hat{R}(t)^{\frac{\hat q-1}{q_2-1}}C(x,t) =k x^{\frac{ \hat q-1}{\min \{q_1,q_2\}-1}}$,
where
$\hat{R}(t)=k_1e^{\lambda_1 t}+k_2\frac{e^{\lambda_2 t}-1}{\lambda_2}$ if $q_1=q_2$,
$k_1^{\frac{q_2-1}{\hat{q}-1}}e^{\frac{q_2-1}{q_1-1}\lambda_1 t}$ if $q_1<q_2$, and
$k_2\frac{e^{\lambda_2 t}-1}{\lambda_2}$ if $q_1>q_2$.
 The proof is the same as that of Theorem \ref{Main-1} with some obvious changes to include $k_i$ and $k$.
}.
\end{theorem}

 Note that only one of $q_i$ is required to be less than $q^*$; the other $q_i$ can be any constant less than one, which means that the  {\it asymptotic marginal behaviour} of  the other utility at large wealth ($U_i'(\infty)$) can be like that of negative power utility ($0<q_i<1$) or log utility ($q_i=0$), not necessarily always  like that of positive power utility ($q_i<0$),  but we still require the value of the utility at zero wealth be zero ($U_i(0)=0$ for $i=1,2$).
Condition  (\ref{V-C-C}) is equivalent to $\lim_{x\to0} U_2'(x)/x^{\hat p-1}=1$ and $\hat p=\hat q/(\hat q-1)>0$, which means  the consumption utility $U_2$ behaves asymptotically like power utility $(1/\hat p)x^{\hat p}$ at small wealth level.
The turnpike property for consumption does not hold  in general,  as the right side of   (\ref{consumption})  is a nonlinear function of $x$ even though the investment horizon is remote, This is because the consumption utility is accumulated over the whole investment period, not just at terminal time,  so one has to choose a consumption policy which achieves a good balance between the initial consumption with small wealth and the future consumption with large wealth.

We next compare our results with those in the literature.
Jin (1998) discusses the turnpike property for a class of utilities $U_i$ such that $(U_i')^{-1}=-V_i'$ are regularly varying at zero with index $1-q_i$, that is,
\footnote{\label{rk4.11}
It is easy to verify that both (\ref{eqn10})  and (\ref{V-wC0}) imply (\ref{R-V}), so  (\ref{R-V}) is the weakest condition among three conditions.
The function $V'(y)=y^{q-1}\ln y (y\leq Y<1)$ is an example that satisfies (\ref{V-wC0}), but not (\ref{eqn10}) (see Huberman and Ross (1983),  page 1346).
 The function  $V'(y)=-y^{q-1}e^{hy\sin\frac{1}{y}}$,
where $h<\frac{1-q}{2}$,   is an example that satisfies (\ref{eqn10}), but not (\ref{V-wC0})  (see 
Back et al. (1999), page 178).
 However, if $\lim_{y\to0} V_i''(y)/y^{q_i-2}$ exists, then both (\ref{eqn10}) and  (\ref{V-wC0}) are satisfied  with  L'Hospital's Rule.
}
\begin{equation}\label{R-V}
 \lim_{y\to 0} {V_i'(xy)\over V_i'(y)} = x^{q_i-1}, \quad\forall x>0.
\end{equation}
  Apart from requiring $(U_i')^{-1}$  to be regularly varying at zero, Jin (1998) also asks the existence of finite limits of $(U_i')^{-1}(x)/x^{q_i-1}$ as $x$ tends to 0 (see Jin (1998, page 1007)), which implies utility functions in Jin (1998) also satisfy  (\ref{eqn10}). We also need condition  (\ref{V-C-C}) for the asymptotic property of optimal consumption policy. There is no explicit counterpart  (\ref{V-C-C}) in Jin (1998), but there is a hidden  assumption in  Jin (1998) (the finiteness of $M_{1,\epsilon}$ and $M_{2,\epsilon}$, toward the bottom of page 1011, which was first pointed out in Back et al.~(1999,  page 194)).  Condition (\ref{V-C-C})  implies $M_{1,\epsilon}$ is finite, but the finiteness of $M_{2,\epsilon}$ is an additional condition in Jin (1998), so these conditions  are not directly comparable.

   Jin (1998, Theorems 5.1-5.3) states that the optimal  portfolio and  consumption strategies of $V_i$ at any fixed time can be approximated by  those of power utilities  if the investment horizon is remote  and the error can be made arbitrarily small in the  absolute or mean-squared norm. The results of Theorem \ref{Main-1} are different. Firstly, we prove the pointwise convergence in (\ref{turnpike}) and (\ref{consumption})  for every fixed $x$,  which is in line with the standard definition of the turnpike property in the literature but  does not imply   norm convergence in the probability space
    as discussed in Jin (1998).  Secondly, we show the turnpike property holds whenever $q_1<q^*$ or $q_2\leq q^*$    whereas  Jin (1998, Theorems 5.4-5.5) requires $q_1=q_2<q^*=0$ ($\delta=0$ in Jin (1998) ) to get the same property.

Back et al. (1999) discuss turnpike property of optimal terminal wealth.
 They claim there is no turnpike property for consumption with a counter example using a translated power utility, see  Back et al. (1999, Section 1.1). However, that utility is $-\infty$  in [0,K) and does not satisfy our assumption of $U_2(0)=0$ and condition (\ref{V-C-C}). Therefore,   Back et al. (1999) only discuss the turnpike property for terminal-wealth utility maximization, different from the one for consumption and terminal-wealth utility maximization  in this paper.

 We now estimate the rate of convergence of the turnpike property if utilities converge to power utilities at certain speed.  The next theorem  establishes the rate of convergence  when $q_1<q^*$ or  $q_2 \leq q^*$,
 which, to the best of our knowledge, is absent in the literature on the turnpike property with consumption and investment.

\begin{theorem} \label{Main-1a}
Assume that $V_i\in C^1(R_+)$ and there are constants $\bar{q}\in (\max\{q_1, q_2,0\}, 1)$
 and $L>0$, such that
\begin{equation}\label{V-C-3}
\left|V'_i(y)+y^{q_i-1}\right|\leq L y^{\bar{q}-1},\ y\leq 1.
\end{equation}
Then $A(x,t)$ converges to $\frac{\theta}{\sigma} (1-\min\{q_1,q_2\})x$ exponentially
\footnote{
Assume $f$ and $g$ are well defined functions on $R_+\times R_+$ and $R_+$, respectively. We say $f(x,t)$ converges to $g(x)$ exponentially (or polynomially) as $t\to \infty$ if there exist constants $c>0$ and $\bar T>0$ and a well defined function $D$ on $R_+$, such that
$|f(x,t)-g(x)|\leq D(x)e^{-ct}$ (or $|f(x,t)-g(x)|\leq D(x)t^{-c}$) for all $x\in R_+$ and $t\geq \bar T$.
Exponential convergence is much faster than polynomial convergence.
}
 as $t\to\infty$ if $q_1<q^*$ or  $q_2 < q^*$, and polynomially if $q_1\geq q^*$ and $q_2=q^*$.
Furthermore, if  there is a constant $\tilde{q}<\min\{\hat q, q_2\}$, such that
\begin{equation}\label{V-C-C-C}
|V'_2(y)+y^{\hat q-1}|\leq L y^{\tilde{q}-1}, y\geq 1,
\end{equation}
then,  $R(t)^{\frac{\hat q-1}{q_2-1}}C(x,t)$ converges to $x^{\frac{\hat q -1}{\min\{q_1,q_2\}-1}}$ exponentially as $t\to\infty$ if $q_1<q^*$ or  $q_2 < q^*$, and polynomially if $q_1\geq q^*$ and $q_2=q^*$, where $R(t)$ is defined in (\ref{R(t)}).

\end{theorem}


\begin{theorem} \label{Main-2}
Let $q_1\geq q^*$ and $q_2> q^*$.
Assume that $V_i\in C^1(R_+)$ and satisfies (\ref{eqn10}).
Then for any $x\in R_+$, we have
\begin{equation}\label{turnpike-2}
\lim_{t\rightarrow \infty}A(x,t)=\frac{\theta}{\sigma }[(1-q_1)Y^{q_1-1}1_{\{q_1=q^*\}}-Yh'(Y)],
\end{equation}
and
\begin{equation}\label{consumption-2}
\lim_{t\rightarrow \infty}C(x,t) =- V'_2(Y),
\end{equation}
where $1_{\{q_1=q^*\}}$ is an indicator which equals 1 if $q_1=q^*$ and 0 otherwise, $Y$ is the unique solution to the equation
\begin{equation}\label{root}
x=y^{q_1-1}1_{\{q_1=q^*\}}+h(y),
\end{equation}
and
\[
h(y)=-\int_{0}^\infty \frac{e^{\beta \tau}}{2\sqrt{\pi }}d\tau\int_{-\infty}^\infty
e^{-\frac{\eta^2}{4}-(\alpha-1) a\sqrt{\tau}\eta}V_2'(ye^{a\sqrt{\tau}\eta})d\eta.
\]
 Furthermore, if $V_2$ is the dual function of a power utility, given by $V_2'(y)=-y^{q_2-1}$ with $q_2<0$, and $q_1, q_2>q^*$, then we have the turnpike property
\begin{equation} \label{eqn2.18}
\lim_{t\to\infty} A(x,t)=
\frac{\theta}{\sigma}(1-q_2)x, \quad
\lim_{t\to\infty} C(x,t)=- \lambda_2 x.
\end{equation}
\end{theorem}

Note that $q_1,q_2<1$ covers all utilities which asymptotically behave like a power utility (including log and negative power utility).
 It is clear from (\ref{turnpike-2}), (\ref{consumption-2}) and (\ref{root}) that the turnpike property in the classical sense does not hold in general when $q_1\geq q^*$ and $q_2>q^*$.
Since $q^*$ tends to $-\infty$ as $\delta$ tends to $\infty$, we have $q_1,q_2>q^*$ when $\delta$ is sufficiently large and the optimal amount of investment in the risky asset is in general a nonlinear function of the wealth even if the investment horizon is very long. This is not surprising economically  as a large discount rate $\delta$ means the utility from terminal wealth in the distant future is negligible and one weighs more on the  utility in the near term from consumption, which then clearly depends on the current wealth level.  This phenomenon  presents only when there are both terminal and consumption utilities. If there is only the terminal utility, then one should still have the turnpike property in the classical sense. We see that the turnpike property with terminal and consumption utilities is fundamentally different from that with terminal utility only.

The next two theorems list some other conditions on utilities that ensure the turnpike property for investment, see Footnote \ref{rk4.11} for relations of different sufficient conditions.

\begin{theorem} \label{Main-3}
Let $q_1, q_2 <q^*$. Assume that $V_i\in C^2(R_+)$ and  satisfy (\ref{V-wC0}).
Assume further that
there are constants $\bar{q}\in (\max\{q_1,q_2,0\}, 1)$, $K>0$ such that\footnote{
If $V_2\equiv 0$, we need only $\bar{q}\in (\max\{q_1, 0\}, 1]$.
}.
\begin{equation}\label{V-wC}
R_i(y)|V_i'(y)|= yV_i''(y) \leq  K y^{\bar{q}-1}, \   y \geq 1.
\end{equation}
Then for any $x\in R_+$, we have the turnpike property (\ref{turnpike})
\end{theorem}

Based on the observation: If $V_i'(y)$ satisfies (\ref{R-V}), then for any $t_0>0$, $v_y(y,t_0)$ satisfies (\ref{V-wC0}), we have
\begin{theorem}\label{Main-4}
Let $q_1, q_2 <q^*$. Assume that $V_1\in C^1(R_+)$ satisfies (\ref{R-V}) with index $1-q_1$,
$V_2\in C^2(R_+)$ satisfies (\ref{V-wC0}) with index $1-q_2$ and (\ref{V-wC}) and $\lim_{y\to 0} {V_2'(y)\over V'_1(y)} = k$ for some $k\in[0,\infty]$ if $q_1 = q_2$.
Then for any $x\in R_+$, we have the turnpike property (\ref{turnpike}).

\end{theorem}

 If $V_2\equiv 0$, Theorem \ref{Main-4} is Theorem 2 in Back et al. (1999). It shows that for  the turnpike property to hold in the presence of consumption,  the terminal wealth utility $U_1$ only needs to satisfy a weaker condition (\ref{R-V}), whereas the consumption utility $U_2$ is required to satisfy a stronger condition  (\ref{V-wC0}).


\section{Examples and numerical tests}\label{Examples}

 In this section we  discuss two examples to illustrate the applications of Theorems \ref{Main-1} and \ref{Main-2}. We will show that if the utility for consumption $U_2$ is not a power utility, then one in general  does not  have the turnpike property when $q_1>q^*$ and $q_2>q^*$, which is in sharp contrast with the result  when $U_2$ is a power utility, see Theorem \ref{Example-1}.

 For $0<p<1$, define a non-HARA utility function
\begin{equation}\label{nonHara}
U(x) = -\frac{1}{q}H(x)^{q} - \frac{1}{\bar{q}}H(x)^{\bar{q}} + xH(x)
\end{equation}
for $x > 0$, where
$H(x) = 2^{1-\bar{p}}\left(\sqrt{1 + 4x}-1\right)^{ \bar{p}-1}$
and $\bar{p} = 2p - 1$, $q = \frac{p}{p-1}$, $\bar{q} = \frac{\bar{p}}{\bar{p}-1}$.
\footnote{We have $\bar q={1\over 2} (q+1)>q$ and $\bar q-1={1\over 2}(q-1)$.
It is easy to verify that $U$ is strictly increasing and strictly concave,    $U(0)=0$, $U(\infty)=\infty$, $U'(0)=\infty$ and $U'(\infty)=0$. Therefore $U$ is a utility function in the classical sense.
This utility (for $p=3/4$) is used in Bian and Zheng (2015) to illustrate the turnpike property and the convergence rate for a terminal wealth utility maximization problem.}
The dual function of $U$ has a  simple form, given by
$V(y) = -{1\over q} y^{q}-{1\over \bar{q}}y^{\bar{q}}$.

\begin{example} \label{ex1}
Assume $U_1$ is a non-HARA utility,  given by (\ref{nonHara}) with $p=p_1$, and $U_2$ is a power utility, given by $U_2'(x)=x^{p_2-1}$,  where $0<p_1,p_2<1$,  which corresponds to $q_1,q_2<0$. Conditions (\ref{eqn10}) and (\ref{V-C-C}) are satisfied with $\hat q=q_2$.  If $q_1<q^*$ or $q_2 \leq q^*$, then we have the turnpike property for investment  (\ref{turnpike}) and the limiting property for consumption (\ref{consumption}). If $q_1> q^*$ and $q_2>q^*$, then we still have the turnpike property for investment  (\ref{eqn2.18}) thanks to the consumption utility $U_2$ being a power utility.
\end{example}

 We now do some numerical tests.  The data used are the same as those in Table  \ref{power_table}.
If we choose $q_1=q_2=q=-3$, which corresponds to $p=3/4$, then $\bar q =-1$ for the non-HARA utility in (\ref{nonHara}).
 We have $\lambda=\lambda(q)=0.16+r$ and $\bar\lambda =\lambda(\bar q)
=0$.  Since $q_1=q_2=q<q^*$, we have  the turnpike property
$$ \lim_{t\to\infty} \pi^*(x,t)= \lim_{t\to\infty} A(x,t) /x = {\theta\over \sigma}(1-q)=4$$
for any fixed wealth level $x$. We may compute the exact optimal trading strategies $\pi^*(x,t)$ and $c^*(x,t)$ to see the accuracy of this approximation. To this end, we need to find the unique solution $y$ to the equation $-v_y(y,t)=x$, that is, $ R(t)y^{q-1} + e^{\bar\lambda t} y^{\bar q-1} = x$,  where $R(t)=e^{\lambda t} + (e^{\lambda t} -1)/\lambda$.
We can solve the equation and get
$$
\pi^*(x,t) =  {1\over x}\left(4R(t) z^2 + 2 z\right)\quad\mbox{and}\quad
c^*(x,t) =  z^2,
$$
 where $z=(-1 + \sqrt{1+ 4xR(t)})/(2R(t))$.
The Merton strategy is given by $\pi_M(x,t):=(\theta/\sigma)(1-q)=4$.
Table  \ref{table1} lists values $\pi^*(x,t)$  (rows 2 to 4) and relative errors to the Merton strategy
$e_M(x,t):={\pi_M(x,t)\over \pi^*(x,t)}-1$   (rows 5 to 7) for $x=10$ and various $r$ and $t$.

It is clear that the optimal investment strategy $\pi^*(x,t)$ converges to the Merton strategy $\pi_M(x,t)$ as $t$ tends to $\infty$, as expected from Theorem \ref{Main-1}. Relative errors $e(x,t)$ show the extent of over-investment (if $e(x,t)>0$) or under-investment (if $e(x,t)<0$) if one takes the Merton strategy $\pi_M(x,t)$ instead of the optimal strategy $\pi^*(x,t)$. For example, for $t=1$ and $r=0.02$, if one takes the Merton strategy, one has over-invested about 10\% compared with the optimal strategy $\pi^*(x,t)$. If the investment period is 25 years or longer, the Merton strategy would produce a relative error of 1\% or less with the data used in this example. The shorter the investment period, the bigger the relative error, up to 10\% for a one-year investment.

\begin{table}
\begin{center}
\begin{tabular}{|c|c|ccccccc|}
\hline
 &$r\backslash t$ & 1&2&5&10&25&50&100\\ \hline
optimal &0.02 &3.6237&	3.7023&	3.8147&	3.8946&	3.9740	&3.9973&	4.0000\\
portfolio  &0.06& 3.6288&	3.7099&	3.8264&	3.9084&	3.9829&	3.9989	 &4.0000\\
$\pi^*(x,t)$ &0.10&3.6340	&3.7175	&3.8376	&3.9209&	3.9889	&3.9996	&4.0000\\
\hline
relative &0.02 &0.1039	&0.0804	&0.0486	&0.0271	&0.0065&	0.0007&	0.0000\\
error&0.06&0.1023	&0.0782&	0.0454&	0.0234	&0.0043	&0.0003	&0.0000\\
 $e(x,t)$ &0.10& 0.1007	&0.0760&	0.0423	&0.0202	&0.0028	&0.0001	&0.0000\\
\hline
\end{tabular}
\end{center}
\caption{Optimal portfolios $\pi^*(x,t)$ and relative errors to the limiting (Merton) portfolio  $e(x,t)$ with different time horizons $t$ and interest rates $r$. The Merton portfolio $\pi_M(x,t)=4$ for all $x$ and $t$.}\label{table1}
\end{table}

Table  \ref{table2} lists values $c^*(x,t)$  (rows 2 to 4), $R(t)c^*(x,t)$  (rows 5 to 7) and relative errors
$f(x,t):={x\over R(t)c^*(x,t)} -1$  (rows 8 to 10) for $x=10$ and various $r$ and $t$.
 It is clear that $\lim_{t\to\infty}R(t)c^*(x,t)=x$, as expected from Theorem \ref{Main-1}. The longer the investment period, the smaller the consumption rate $c^*(x,t)$.  Relative errors $f(x,t)$ show the extent of over-consumption  (if $f(x,t)>0$) or under-consumption  (if $f(x,t)<0$) if one takes the limiting consumption $x$, discounted by $R(t)$, i.e., $c(x,t)=x/R(t)$. For example, for $t=1$ and $r=0.02$, if one takes the consumption  strategy $c(x,t)=$, one has over-consumed about 23\% compared with the optimal consumption strategy $c^*(x,t)$. If the investment period is 25 years or longer, the consumption strategy $c(x,t)$ would produce a relative error of 1\% or less with the data used in this example. The shorter the investment period, the bigger the relative error, up to 23\% for a one year investment.

\begin{table}
\begin{center}
\begin{tabular}{|c|c|ccccccc|}
\hline
 &$r\backslash t$ & 1&2&5&10&25&50&100\\ \hline
optimal&0.02 &3.5407&	2.2161&	0.8585&	0.2778&	0.0169&	0.0002&	0.0000\\
consumption&0.06&3.4442&	2.1033	&0.7538	&0.2097&	0.0073&	0.0000&	0.0000\\
$c^*(x,t)$&0.10&3.3497	&1.9947&	0.6593&	0.1564&	0.0031	&0.0000	&0.0000\\
\hline
&0.02 &8.1183&	8.5113	&9.0734	&9.4730	&9.8701	&9.9863	&9.9998\\
$R(t)c^*(x,t)$&0.06&8.1441&	8.5497	&9.1318&9.5421	&9.9144	&9.9945	&10.0000\\
&0.10&8.1698	&8.5877&	9.1880	&9.6045&	9.9444	&9.9978	&10.0000\\
\hline
relative&0.02&0.2318&	0.1749&	0.1021	&0.0556	&0.0132	&0.0014	&0.0000\\
error&0.06&0.2279	&0.1696	&0.0951	&0.0480&	0.0086	&0.0005	&0.0000\\
$f(x,t)$&0.10&0.2240&	0.1645&	0.0884&	0.0412&	0.0056	&0.0002&	0.0000\\
\hline
\end{tabular}
\end{center}
\caption{Optimal consumptions  $c^*(x,t)$, time adjusted optimal consumptions $R(t)c^*(x,t)$, and relative errors  to the wealth $f(x,t)$ with different time horizons $t$ and interest rates $r$, the wealth level $x=10$.}\label{table2}
\end{table}

Table  \ref{table3} lists absolute errors to the Merton strategy $\bar e(x,t)=|\pi^*(x,t)-\pi_M(x,t)|$ (rows 2 to 4) and approximate convergence rate $c_n$ (rows 5 to 7) for $x=10$ and various $r$ and $t$.~\footnote{If $\{e_n\}$ is a sequence of errors with exponential convergence, then there are positive constants $M$ and $c$ such that $|e_n|\leq Me^{-cn}$. To find $c$, we may assume $|e_n|\approx Me^{-cn}$, which gives $ c\approx -\ln\left(|e_{n+1}|/|e_n|\right)$. In our numerical tests, we have chosen time horizon $t=1,2,5,10,25,50,100$, which does not have equally spaced intervals. An adjustment is needed to reflect this. Specifically, we estimate $c$ by
$ c_n:=- (1/m)\ln \left(|e_{n+m}|/ |e_n|\right)$,
where $m$ is an integer indicating the distance of indices $n$ and $n+m$ for  errors $e_n$ and $e_{n+m}$.
For $n:=1,2,5,10,25,50,100$, the distance between adjacent points are
$m=1,3,5,15,25,50$. We form a sequence $\{c_n\}$ to see if there is a limit which would indicate the approximate exponent of the exponential convergence rate.
}
It is clear from Table  \ref{table3} that there is convergence of the error sequence for fixed $r$.  For $r=0.02$, 0.06, and 0.10, the exponent of exponential convergence rate is approximately equal to $c=0.09$, 0.11, and 0.13, respectively.

\begin{table}
\begin{center}
\begin{tabular}{|c|c|ccccccc|}
\hline
 &$r\backslash t$ & 1&2&5&10&25&50&100\\ \hline
absolute &0.02 &0.3763&	0.2977&	0.1853&	0.1054&	0.0260&	0.0027&	0.0000\\
error &0.06&0.3712	&0.2901	&0.1736	&0.0916	&0.0171	&0.0011	&0.0000\\
$\bar e(x,t)$ &0.10& 0.3660	&0.2825&0.1624&	0.0791&	0.0111	&0.0004&	0.0000\\
\hline
convergence &0.02 &&0.2343&	0.1580	&0.1128&	0.0933&	0.0900&	0.0900\\
rate &0.06&&0.2466&	0.1710&	0.1279	&0.1118	&0.1099	&0.1100\\
$c_n$&0.10&&0.2592	&0.1845	&0.1439	&0.1308&	0.1299	&0.1300
\\
\hline
\end{tabular}
\end{center}
\caption{Absolute errors between the optimal portfolios and the Merton portfolio $\bar e(x,t)$, distance between adjacent points $m$, and estimations of rate of convergence $c_n$ with different time horizons $t$ and interest rates $r$.} \label{table3}
\end{table}

We have done the same set of numerical tests with  $q_1=q_2=q=-1/3$ and all other data the same as above, which corresponds to $p=1/4$ and $\bar q =1/3$ for the non-HARA utility in (\ref{nonHara}).  Since $q_1=q_2=q>q^*$, we have
$\lambda<0$  and
$ \lim_{t\to\infty}R(t)=-1/ \lambda$. Numerical tests show that  $\lim_{t\to\infty} \pi^*(x,t)=(\theta/\sigma)(1-q)$ and $\lim_{t\to\infty} c^*(x,t)=-\lambda x$. Therefore, the turnpike property for investment still holds due to the consumption utility being a power utility. The numerical tests also indicate that the convergence is exponential although we have not proved this result when $q_1$ and $q_2$ are greater than $q^*$.

\begin{example}
Assume $U_1$ is a power utility, given by $U_1'(x)=x^{p_1-1}$, and  $U_2$ is a non-HARA utility, given by (\ref{nonHara}) with $p=p_2$, where $0<p_1,p_2<1$. Condition (\ref{eqn10}) is satisfied.

 If $q_1<q^*$ or  $q_2 \leq q^*$, then from (\ref{turnpike}), we have
$$\lim_{t\rightarrow \infty}A(x,t)=\frac{\theta}{\sigma}(1-\min\{q_1,q_2\})x.
$$

If $q_1> q^*$ and $q_2>q^*$, then  from  (\ref{turnpike-2}) and (\ref{consumption-2}), we have
$$
\lim_{t\rightarrow \infty}A(x,t)=-\frac{\theta}{\sigma } Yh'(Y),\quad
\lim_{t\rightarrow \infty}C(x,t) =-V_2'(Y),
$$
where  $Y$ is the unique solution to the equation
$x=- (1/\lambda_2) y^{q_2-1} - (1/\bar\lambda_2) y^{\bar q_2-1}$.
Solving the equation  and substituting $Y$ into the limits  above, we get
\begin{equation} \label{A_C_ex2}
\lim_{t\rightarrow \infty}A(x,t)=\frac{\theta}{\sigma }(1-q_2)\left(
 x + {1\over 2\bar \lambda_2}  Z \right),\quad
\lim_{t\rightarrow \infty}C(x,t) =Z^2 + Z,
\end{equation}
 where $Z=  2x\left(\sqrt{\frac{1}{\bar{\lambda}_2^2}-\frac{4x}{\lambda_2}}-\frac{1}{\bar{\lambda}_2}\right)^{-1}$.

Clearly, the limiting optimal portfolio and  consumption are not linear functions of  wealth,  in other words, the initial wealth level would affect the behaviour of optimal trading strategies even though the time horizon is very long. The turnpike property in the classical sense does not hold in this scenario. This is in a marked contrast to the case when the consumption utility is a power utility function. We may conclude that there is no turnpike property in general when the consumption utility is a general utility.
\end{example}

 We have done numerical tests with the same data as in Example \ref{ex1}. When $q_1=q_2=q=-3<q^*=-1$, we still have the turnpike property for investment.

When $q_1=q_2=q=-1/3>q^*=-1$ and $\bar q_2=\bar q=1/3$, we know there is no turnpike property in general. To illustrate this point numerically, we find the exact optimal investment and consumption strategies $\pi^*(x,t)$ and $c^*(x,t)$ as follows: find the unique solution $y$ to the equation $-v_y(y,t)=x$, that is,
 $R(t) y^{q-1} +  R_1(t) y^{\bar q-1}=x$, where $R(t)=e^{\lambda t} + (e^{\lambda t} -1)/ \lambda$ and
$R_1(t)= (e^{\bar\lambda t} -1)/ \bar\lambda$. We can solve the equation and get
$$
\pi^*(x,t) =  {1\over x}\left({4\over 3}R(t) z^2 + {2\over 3}  R_1(t)z\right)\quad\mbox{and}\quad
c^*(x,t) =  z^2+z,
$$
where $z=(-R_1(t)+\sqrt{R_1(t)^2+4xR(t)}/(2R(t))$.
Using (\ref{A_C_ex2}), also noting $\theta/\sigma=1$ and $q_2=-1/3$, we can find relative errors
$$ e(x,t)= {{4\over 3} (1+{1\over 2\bar\lambda x}Z)\over \pi^*(x,t)}-1
\quad\mbox{and}\quad
f(x,t)={Z^2+Z\over c^*(x,t)}-1.$$
The Merton strategy is given by $\pi_M(x,t)=(\theta/\sigma)(1-q)=4/3$.

Table  \ref{table4}  lists values $\pi^*(x,t)$  (rows 2 to 4), relative errors
$e(x,t)$  (rows 5 to 7), and relative errors to the Merton strategy $e_M(x,t)$ (rows 8 to 10) for $x=10$ and various $r$ and $t$.
It is clear that $\pi^*(x,t)$ does not converge to the Merton strategy  as $t$ tends to $\infty$. If one used the Merton strategy, one would greatly over-invest the risky asset compared with the optimal investment strategy $\pi^*(x,t)$. This phenomenon is due to the consumption utility not being a power utility.

\begin{table}
\begin{center}
\begin{tabular}{|c|c|ccccccc|}
\hline
 &$r\backslash t$ & 1&2&5&10&25&50&100\\ \hline
optimal &0.02 &1.2008	&1.1328&	1.0290	&0.9519	&0.8710&	0.8353	&0.8222\\
portfolio  &0.06& 1.2017&	1.1356&	1.0386&	0.9722&	0.9151&	0.9025&	0.9055\\
$\pi^*(x,t)$ &0.10&1.2027&	1.1384&	1.0478&	0.9912&	0.9538&	0.9540&	0.9588\\
\hline
relative &0.02 &-0.3155	&-0.2744	&-0.2012	&-0.1365	&-0.0563&	-0.0160	&-0.0003\\
error&0.06&-0.2450&	-0.2011&	-0.1265&	-0.0668&	-0.0086&	0.0053&	0.0020\\
 $e(x,t)$ &0.10& -0.2023&	-0.1573&	-0.0845&	-0.0321	&0.0059&	0.0056	&0.0006
\\
\hline
relative &0.02 &0.1104	&0.1771&	0.2958&	0.4007&	0.5307&	0.5962	&0.6216\\
error&0.06&0.1095&	0.1741	&0.2838&	0.3714	&0.4570	&0.4775&	0.4725\\
 $e_M(x,t)$ &0.10& 0.1086	&0.1712	&0.2724	&0.3451	&0.3980	&0.3976&	0.3907\\
\hline
\end{tabular}
\end{center}
\caption{Optimal portfolios $\pi^*(x,t)$, relative errors to the limiting portfolio $e(x,t)$,  and relative errors to the Merton portfolio $e_M(x,t)$ with different time horizons $t$ and interest rates $r$. The Merton portfolio $\pi_M(x,t)=4/3$ for all $x$ and $t$.}\label{table4}
\end{table}

 Table \ref{table5}  lists values $\pi^*(x,t)$ for different wealth levels $x=1, 10, 100$ and different  time horizons $t$ and interest rates $r$. It is clear the optimal portfolios do not converge to the Merton portfolio and  are dependant of the wealth level $x$ even when the time horizon is very long.

\begin{table}
\begin{center}
\begin{tabular}{|c|c|ccccccc|}
\hline
$x$ &$r\backslash t$ & 1&2&5&10&25&50&100\\ \hline
 &0.02 &1.0014&	0.8922&	0.7845&	0.7365&	0.7045&	0.6944&	0.6912\\
1  &0.06& 1.0032&	0.8961&	0.7921&	0.7472&	0.7201&	0.7152	&0.7163\\
 &0.10&1.0051	&0.8999	&0.7998	&0.7583	&0.7374&	0.7375&	0.7400\\
\hline
 &0.02 &1.2008	&1.1328&	1.0290	&0.9519	&0.8710&	0.8353	&0.8222\\
10  &0.06& 1.2017&	1.1356&	1.0386&	0.9722&	0.9151&	0.9025&	0.9055\\
&0.10&1.2027&	1.1384&	1.0478&	0.9912&	0.9538&	0.9540&	0.9588\\
\hline
 &0.02 &1.2881&	1.2617&	1.2149&	1.1727&	1.1165&	1.0852&	1.0723\\
100  &0.06& 1.2885&	1.2628&	1.2197&	1.1847&	1.1491	&1.1402	&1.1424\\
 &0.10&1.2888	&1.2640	&1.2242&	1.1953	&1.1738&	1.1739&	1.1768\\
\hline
\end{tabular}
\end{center}
\caption{Optimal portfolios $\pi^*(x,t)$ with different wealth levels $x$, time horizons $t$, and interest rates $r$. The Merton portfolio $\pi_M(x,t)=4/3$ for all $x$ and $t$.}\label{table5}
\end{table}

\section{Conclusions}
 In this paper we discuss the turnpike property for optimal investment and consumption problems. We use the dual control method to characterize the optimal investment and consumption strategies in terms of the dual value function. We find there exists a threshold value that determines if the turnpike property for investment holds when utilities from consumption and terminal wealth behave like power utilities at large wealth. We show the turnpike property for consumption does not hold in general. We derive the exponential rate of   convergence of the optimal strategies to their limiting strategies.  We  illustrate the main results with two examples using power and non-HARA utilities and some numerical tests.

\bigskip
{\noindent\bf Acknowledgment}. The authors are grateful to the anonymous reviewer whose comments and suggestions  have helped to improve the previous two versions.


\section{Appendix} \label{proofs}

\noindent{\bf Derivation of optimal trading strategy (\ref{ICS-1}) and budget equation (\ref{bce})}.
We may use a standard stochastic control method to solve problem (\ref{primal_problem}). Define the value function $u$  by
$$ u(x,t)=\sup_{\pi, c} E\left[\int_t^T e^{-\delta (s-t)} U_2(c_s)ds +e^{-\delta (T-t)}U_1(X_T)\bigg | X_t=x\right]
$$
for $(x,t)\in R_+\times [0,T]$. Then $u$  satisfies the  HJB equation
\begin{equation} \label{HJBE-1}
{\partial u\over \partial t}+\max_{\pi} \{ \frac{1}{2}\sigma^2x^2\pi^2 u_{xx}+x\pi\sigma \theta u_x\}+rxu_x-\delta u+\max_{c\geq 0}\{U_2(c)-cu_x\}=0,
\end{equation}
for $(x,t)\in R_+\times [0,T)$ with the terminal condition $u(x,T)=U_1(x)$
for $x\in R_+$, where $u_x(x,t)$ is the partial derivative of $u$ with respect to $x$ and evaluated at $(x,t)$ (we have omitted  $(x,t)$ in equation (\ref{HJBE-1}) to simplify notations), ${\partial\over \partial t}u$ and $u_{xx}$ are defined similarly.

 Let $V_i$ be the dual functions of $U_i$, $i=1,2$, defined in (\ref{dual_func_V_i}).
Then $V_i$ are nonnegative (since $U_i(0)=0$), continuously differentiable, strictly decreasing, strictly convex functions.

The optimal investment and consumption strategies in the HJB equation (\ref{HJBE-1}) are given by
\begin{equation} \label{ICS-10}
\pi^*(x,t)=- \frac{\theta}{\sigma}\frac{u_x}{xu_{xx}},\ \ \ c^*(x,t)=-V'_2 (u_x).
\end{equation}
The HJB equation (\ref{HJBE-1}) can be written as
\begin{equation} \label{HJBE-1a}
{\partial u\over \partial t}-\frac{1}{2}\theta^2{u_x^2 \over u_{xx}}+rxu_x-\delta u+V_2(u_x)=0.
\end{equation}
Equation (\ref{HJBE-1a}) is a fully nonlinear PDE and is in general difficult to solve. In the literature equation (\ref{HJBE-1a}) is solved with some trial-and-error method, which works when $U_1$ and $U_2$ are the same power utility functions. Bian et al. (2011) and Bian and Zheng (2015)  apply the  dual stochastic control method to solve problem (\ref{primal_problem}) and show that the HJB equation (\ref{HJBE-1}) has a classical solution that can be represented in terms of the dual function of the corresponding dual control problem. The value function $v(y,t)$ of the dual control problem satisfies a  linear PDE
 \begin{equation} \label{dual_HJB}
{\partial v\over \partial t} + {1\over 2}\theta^2 y^2 v_{yy}
-(r-\delta)yv_y - \delta v  +V_2(y) =0
\end{equation}
for $(y,t)\in R_+\times [0,T)$, with the terminal condition
$v(y,T)=V_1(y)$.
 Its solution $v$ is continuous on  $R_+\times [0,T]$ and
$C^{2,1}$ on $R_+\times [0,T)$, and $v$ is strictly decreasing and strictly convex in $y$ for fixed $t<T$. A classical solution to the HJB equation (\ref{HJBE-1}) is given by
$$
u(x,t) = v(y, t) + x y,
$$
where $y$ is the unique solution to the equation
 $$
x=-v_y(y,t).
$$
This leads to $u_x(x,t)=y$ and $u_{xx}=-1/v_{yy}$.  From (\ref{ICS-10}) we conclude that the optimal amount of investment $A(x,t)$ and the optimal consumption rate $C(x,t)$  are given by
$$
A(x,t)=x\pi^*(x,t)=\frac{\theta}{\sigma}y v_{yy}(y,t) ,\ \ \  C(x,t)= c^*(x,t)=-V'_2 (y).
$$

 To simplify the notation, we define $\tau=T-t$, the time horizon. Then $T\to \infty$ is equivalent to $\tau\to\infty$. We still use $t$ to represent a time horizon variable, instead of $\tau$.

It is easy to verify that $v_y(y,t)$
is the solution of the initial value problem
\begin{equation}\label{x}
w_t-\frac{1}{2}\theta^2 y^2 w_{yy}+(r-\delta-\theta^2)yw_y+ \delta w=V_2^{'}(y)
\end{equation}
for $(y,t)\in R_+\times R_+$ with  initial condition $v_y(y,0)=V'_1(y),\ y\in R_+$.
By Poisson's formula, the solution of equation (\ref{x}) is given by
\begin{equation}\label{PF-2}
v_y(y,t)=-I_1(y,t)-\int_{0}^tI_2(y,\tau)d\tau,\ yv_{yy}(y,t)=J_1(y,t)+\int_{0}^tJ_2(y,\tau)d\tau,
\end{equation}
 which proves  (\ref{ICS-1})  and (\ref{bce}).

\bigskip\noindent
{\bf Preliminaries of mathematical proofs}.
We need some technical results in proofs.  Simple calculus gives that for any constant $A$,
\begin{equation}\label{A}
\frac{1}{2 \sqrt{\pi }}\int_{-\infty}^{\infty}
e^{-\frac{\eta^2}{4}-A\eta}d\eta=e^{A^2 }
\end{equation}
Hence for any $q<1$,
\begin{equation}\label{C-1}
\frac{e^{\beta t}}{2\sqrt{\pi }}\int_{-\infty}^\infty
e^{-\frac{\eta^2}{4}-(\alpha-1) a\sqrt{t}\eta}(\frac{\eta}{2a\sqrt{t}}+\alpha-1)e^{(q-1)a\sqrt{t}\eta}d\eta
=(q-1)e^{\lambda(q) t}
\end{equation}
and
\begin{equation}\label{C-2}
\frac{e^{\beta t}}{2\sqrt{\pi }}\int_{-\infty}^\infty
e^{-\frac{\eta^2}{4}-(\alpha-1) a\sqrt{t}\eta}|\frac{\eta}{2a\sqrt{t}}+\alpha-1|e^{(q-1)a\sqrt{t}\eta}d\eta
\leq [1-q+\frac{1}{a\sqrt{\pi t}}]e^{\lambda(q) t}.
\end{equation}

For any $y>0$, the convexity of $V_i$ implies that
\[
 V_i(\frac{y}{2})\geq V_i(y)-\frac{y}{2} V_i'(y),
\]
which, together with the decreasing property and the nonnegativity of $V$, gives
\begin{equation} \label{V-NC}
0\leq -yV_i'(y) \leq 2 V_i(\frac{y}{2}) \leq 2V_i(\frac{1}{2}),
\end{equation}
for all $y\geq 1$.

We need the following algebraic inequality in the proof:
\begin{equation}\label{e51}
|x^c-y^c|\leq (1+c)\max\{x, y\}^{c-1}|x-y|
\end{equation}
for all $x,y\geq 0$ and $0<c<\infty$. (\ref{e51}) can be proved as follows: assume $x>y>0$ and let $z=x/y$. Then $z>1$ and (\ref{e51}) is equivalent to $z^c -1 \leq (1+c)(z^c-z^{c-1})$. Let $g(z)=z^c -1-(1+c)(z^c-z^{c-1})$. Then  $g'(z)=cz^{c-1}-(1+c)(cz^{c-1}-(c-1)z^{c-2})=-c^2(1-z)z^{c-2}-z^{c-2}\leq 0$ for $z\geq 1$. So $g$ is decreasing and $g(z)\leq g(1)=0$ for $z\geq 1$.

We have, for any $q<\bar{q}<1$, $\lambda=\lambda(q), \bar{\lambda}=\lambda(\bar{q})$
\begin{equation}\label{1}
e^{\bar{\lambda} t}y^{\bar{q}-1}=(e^{\lambda t}y^{q-1})^{\frac{\bar{q}-1}{q-1}}e^{(\bar{\lambda}-\lambda\frac{\bar{q}-1}{q-1}) t}.
\end{equation}
and
\begin{equation}\label{2}
\bar{\lambda}-\lambda\frac{\bar{q}-1}{q-1}=(\bar{q}-q)[\frac{1}{2}\theta^2(\bar{q}-1)+\frac{r}{q-1}]<0.
\end{equation}

\bigskip\noindent
{\bf Proof of Theorem \ref{Example-1}}.
Since $U'_i(x)= x^{p_i-1}$ for $i=1,2$, the dual functions are given by $V'_i(y)=-y^{q_i-1}$ where $q_i=p_i/(p_i-1)$.
Substituting $V'_i(y)$  into (\ref{PF-2}), also noting (\ref{A}), we have
$$ v_y(y,t)=-e^{\lambda_1 t} y^{q_1-1} -\frac{e^{\lambda_2 t }-1}{\lambda_2} y^{q_2-1}.$$
Hence we derive (\ref{control-i}) from (\ref{ICS-1}),
where $y$ is the solution to the equation (\ref{relationship}).

We next complete the proof by directly discussing three cases.

{\it Case 1}. When $q_1<q^*$ or $q_2\leq q^*$, we have $\lambda_1>0$ or $\lambda_2\geq 0$. We know that the solution $y=u_x$ to equation (\ref{relationship}) must tend to $\infty$ as $t$ tends to $\infty$ (otherwise the right side of  (\ref{relationship}) tends to $\infty$).
We can find the dominating term in (\ref{relationship}) as $t$ tends $\infty$ for different $q_1$ and $q_2$.

Assume $q=q_1=q_2 \leq q^*$. It is easy to see from (\ref{ICS-1}) and (\ref{relationship}) that $A(x,t)=\frac{\theta}{\sigma} (1-q)x$ and
$C(x,t)=-V'_2 (u_x(x,t))=(u_x(x,t))^{q_2-1}=E(t)x$, where $E(t)=(e^{\lambda_1 t}  +\frac{e^{\lambda_2 t }-1}{\lambda_2})^{-1}$.  This implies $R(t)C(x,t)=x$.

Assume $q_1<\min\{q_2, q^*\}$, then $\lambda_1>\max\{\lambda_2, 0\}$. Noting that
\[
\frac{e^{\lambda_2 t }-1}{\lambda_2}=\frac{1-e^{-|\lambda_2| t}}{|\lambda_2|} e^{\lambda^+_2 t}\leq te^{\lambda^+_2 t},
\]
we obtain
\begin{equation} \label{i-c}
\frac{e^{\lambda_2 t }-1}{\lambda_2}y^{q_2-1}\leq (e^{\lambda_1 t}y^{q_1-1})^{\frac{q_2-1}{q_1-1}}te^{(\lambda^+_2-\lambda_1\frac{q_2-1}{q_1-1}) t}
\leq x^{\frac{q_2-1}{q_1-1}}te^{(\lambda^+_2-\lambda_1\frac{q_2-1}{q_1-1}) t}.
\end{equation}
From (\ref{relationship}) and (\ref{2}), we deduce that
$\lim_{t\rightarrow \infty}\frac{e^{\lambda_2 t }-1}{\lambda_2}y^{q_2-1}=0$ and $\lim_{t\rightarrow \infty}e^{\lambda_1 t}y^{q_1-1}=x$.
Hence from (\ref{control-i})
\[
\lim_{t\rightarrow \infty}A(x,t)=\frac{\theta}{\sigma}(1-q_1)x.
\]
We have by (\ref{control-i})
\[
e^{\lambda_1\frac{q_2-1}{q_1-1} t}C(x,t)=(e^{\lambda_1 t}y^{q_1-1})^{\frac{q_2-1}{q_1-1}},
\]
which implies $\lim_{t\to\infty} R(t) C(x,t)=x^{\frac{q_2-1}{q_1-1}}$.

Assume $q_2< q_1$ and $q_2\leq q^*$, then $\lambda_2> \lambda_1$ and $\lambda_2\geq 0$. We have
\begin{equation} \label{E}
\frac{\lambda_2 }{1-e^{-\lambda_2 t}}\leq E:=\frac{\lambda_2 }{1-e^{-\lambda_2 }}
\end{equation}
for $t\geq 1$. Then by (\ref{relationship}), for $t\geq 1$
\begin{equation} \label{c-i}
e^{\lambda_1 t}y^{q_1-1} = (\frac{e^{\lambda_2 t }-1}{1-e^{-\lambda_2 t}} y^{q_2-1})^{\frac{q_1-1}{q_2-1}} e^{(\lambda_1-\lambda_2\frac{q_1-1}{q_2-1}) t}\leq  (Ex)^{\frac{q_1-1}{q_2-1}} e^{(\lambda_1-\lambda_2\frac{q_1-1}{q_2-1}) t}.
\end{equation}
We deduce from (\ref{relationship}) and (\ref{2}) $\lim_{t\rightarrow \infty}e^{\lambda_1 t}y^{q_1-1}=0$ and
$\lim_{t\rightarrow \infty}\frac{e^{\lambda_2 t }-1}{\lambda_2}y^{q_2-1}=x$.
Furthermore, we get
\[
\lim_{t\rightarrow \infty}A(x,t)=\frac{\theta}{\sigma}(1-q_2)x,
\]
and $\frac{e^{\lambda_2 t }-1}{\lambda_2}C(x,t)= \frac{e^{\lambda_2 t }-1}{\lambda_2}y^{q_2-1}$, which gives $\lim_{t\to\infty} R(t)c(x,t)=x$.

{\it Case 2}. Since $q_1> q^*$  and $q_2>q^*$, we have $\lambda_1, \lambda_2< 0$. From (\ref{relationship}), we deduce that $y=u_x(x,t)$ must have
 upper bound in $t$ (otherwise, we would have $x=0$, a contradiction). Therefore $x=\lim_{t\to\infty}(-1/\lambda_2)u_x(x,t)^{q_2-1}$. Taking the limit in (\ref{control-i})  leads to the desired results.

{\it Case 3}. Since $q_1= q^*$  and $q_2>q^*$, we have $\lambda_1=0$ and $\lambda_2<0$. Again from (\ref{relationship}), we deduce that $y=u_x(x,t)$ must have
upper bound. Let $Y$ be the unique solution to the nonlinear equation
\[
x=Y^{q^*-1}-\frac{1}{\lambda_2}Y^{q_2-1},
\]
then we have $\lim_{t\rightarrow \infty}u_x=Y$. Taking the limit in (\ref{control-i})  leads to the desired results.
\qed

\bigskip\noindent
{\bf Proof of Theorem \ref{Main-1}}.
From (\ref{eqn10}) and (\ref{V-NC}), we conclude that there is constant $\bar{q}\in (\max\{q_1,q_2,0\}, 1)$,
for any fixed $\epsilon>0$, there is $K_\epsilon>0$, such that
\begin{equation} \label{V-C-1a}
|V'_i(y)+ y^{q_i-1}|\leq \epsilon |V'_i(y)| + K_\epsilon y^{\bar{q}-1}.
\end{equation}
for $i=1, 2$. By (\ref{V-C-1a})(i=1) and (\ref{A}), we get
\[
|I_1(y,t)-y^{q_1-1}e^{\lambda_1 t}|\leq \epsilon I_1(y,t)+K_\epsilon y^{\bar{q}-1}e^{\bar{\lambda} t},
\]
for all $(y,t)\in R_+\times (0,\infty)$. Taking $\epsilon=1$ and denoting $\bar{K}=K_1$, we get, by  (\ref{1}) and (\ref{2})
\[
y^{q_1-1}e^{\lambda_1 t}\leq 2 I_1(y,t)+\bar{K} (y^{q_1-1}e^{\lambda_1 t})^{\frac{\bar{q}-1}{q_1-1}}.
\]
 Since $0<\frac{\bar{q}-1}{q_1-1}<1$, we see that $(y^{q_1-1}e^{\lambda_1 t})^\frac{q_1-\bar{q}}{q_1-1} \leq 2 I_1(y,t)+ \bar{K}$
if $y^{q_1-1}e^{\lambda_1 t}\geq 1$. Hence
\[
y^{q_1-1}e^{\lambda_1 t}\leq [2 I_1(y,t)+1+ \bar{K} ]^{\frac{q_1-1}{q_1-\bar{q}}}.
\]
Similarly, we obtain for $t\geq 1$
\[
y^{q_2-1}\frac{e^{\lambda_2 t}-1}{\lambda_2}\leq \left[2 \int_{0}^tI_2(y,\tau)d\tau +1+ \frac{\bar{K}E^{\bar{q}-1\over q_2-1}}{|\bar{\lambda}|} \right]^{\frac{q_2-1}{q_2-\bar{q}}}.
\]
Let $H_1(t)=(u_x(x,t))^{q_1-1}e^{\lambda_1 t}$,
$H_2(t)=(u_x(x,t))^{q_2-1}\frac{e^{\lambda_2 t}-1}{\lambda_2}$, $M_1(t)=(u_x(x,t))^{\bar{q}-1}e^{\bar{\lambda} t}$,
$M_2(t)=(u_x(x,t))^{\bar{q}-1}\frac{e^{\bar{\lambda} t}-1}{\bar{\lambda}}$, $H(t)=H_1(t)+H_2(t)$ and $M(t)=M_1(t)+M_2(t)$.
Then by (\ref{bce})
\begin{equation} \label{V-E-3}
H(t)\leq H:=\left[2x+1+ \bar{K}\right]^{\frac{q_1-1}{q_1-\bar{q}}}+\left[2x +1+ \frac{\bar{K}E^{\bar{q}-1 \over q_2-1}}{|\bar{\lambda}|} \right]^{\frac{q_2-1}{q_2-\bar{q}}}.
\end{equation}

Recalling $E(t)=(e^{\lambda_1 t}+\frac{e^{\lambda_2 t}-1}{\lambda_2})^{-1}$. For $q_1<q^*$ or $q_2 \leq q^*$, we see $\lambda_1>0$ or $\lambda_2 \geq 0$. Hence
$E(t)\leq 1$ for $t\geq 1$ and $\lim_{t\rightarrow \infty}E(t)=0$. We have for all $y>0$,
\[
y^{q_1-1}e^{\lambda_1 t}+y^{q_2-1}\frac{e^{\lambda_2 t}-1}{\lambda_2}\geq E(t)^{-1}\min\{
y^{q_1-1},y^{q_2-1}\} .
\]
Hence
\begin{equation} \label{V-E-y}
u_x(x,t) \geq  \min\{
(E(t)H(t))^{\frac{1}{q_1-1}}, (E(t)H(t))^{\frac{1}{q_2-1}}\}.
\end{equation}
This implies, for $t\geq 1$
\begin{equation}\label{V-E-4}
u_x(x,t)\geq  H^{\frac{1}{\max\{q_1,q_2\}-1}}
(E(t))^{\frac{1}{\min\{q_1,q_2\}-1}},
\end{equation}
and $\lim_{t\rightarrow \infty}u_x(x,t)=\infty$.
Using (\ref{1}) and (\ref{2}), we derive the following estimates:
\begin{equation}\label{V-E-m}
M(t) \leq \tilde{M}(x,t):=H^{\frac{\bar{q}-1}{q_1-1}}e^{(\bar{\lambda}-\lambda_1\frac{\bar{q}-1}{q_1-1}) t}
+ \frac{1}{|\bar{\lambda}|}
H^{\frac{\bar{q}-1}{\max\{q_1,q_2\}-1}}(E(t))^{\frac{\bar{q}-1}{\min\{q_1,q_2\}-1}}.
\end{equation}

By (\ref{V-C-1a}), (\ref{bce}) and (\ref{A}), we get
\begin{equation} \label{V-E-2}
|H(t)-x|\leq \epsilon H(t) + K_\epsilon M(t).
\end{equation}
Noting that (\ref{2}), (\ref{V-E-m}), and letting $t\to\infty$ and then $\epsilon\to 0$ in (\ref{V-E-2}), we deduce the limiting form of (\ref{relationship})
\begin{equation}\label{V-E-5}
\lim_{t\rightarrow \infty}(u_x(x,t))^{q_1-1}e^{\lambda_1 t}+(u_x(x,t))^{q_2-1}\frac{e^{\lambda_2 t}-1}{\lambda_2}=x.
\end{equation}

By (\ref{V-C-1a}),(\ref{ICS-1}), (\ref{C-1}) and (\ref{C-2}), we conclude that
\begin{eqnarray*}
& &\left|yv_{yy}(y,t)-(1-q_1)y^{q_1-1}e^{\lambda_1 t}-(1-q_2)y^{q_2 -1}\frac{e^{\lambda_2 t}-1}{\lambda_2}\right|\\
&\leq&  \epsilon \left[(1-q_1+\frac{1}{a\sqrt{\pi}})y^{q_1-1}e^{\lambda_1 t}+(1-q_2+\frac{2Ee^{\lambda^+_2}+1}{a\sqrt{\pi}})y^{q_2-1}\frac{e^{\lambda_2 t}-1}{\lambda_2}\right]\\
& &{} + K_\epsilon \left[(1-\bar{q}+\frac{1}{a\sqrt{\pi}})y^{\bar{q}-1}e^{\bar{\lambda} t}+(1-\bar{q}+\frac{2Ee^{\lambda^+_2}+1}{a\sqrt{\pi}})y^{\bar{q}-1}\frac{e^{\bar{\lambda} t}-1}{\bar{\lambda}}\right]
\end{eqnarray*}
for all $(y,t)\in R_+\times [1,\infty)$.
Let
\[
\bar{D}=\max\{1-q_1+\frac{1}{a\sqrt{\pi}}, 1-q_2+\frac{2Ee^{\lambda^+_2}+1}{a\sqrt{\pi}}\}.
\]
Noting $q_1,q_2<\bar{q}$, for fixed $x>0$, we obtain
\begin{equation} \label{V-E-6}
\left|A(x,t)-\frac{\theta}{\sigma}[(1-q_1)H_1(t)+(1-q_2)H_2(t)]\right|
\leq
\frac{\theta\bar{D}}{\sigma}(\epsilon H(t) +K_\epsilon M(t)).
\end{equation}
From estimates (\ref{V-E-3}) and (\ref{V-E-m}), letting $t\to\infty$ and then $\epsilon\to 0$ in (\ref{V-E-6}), we derive
\begin{equation}\label{V-E-7}
\lim_{t\rightarrow \infty}\left[A(x,t)-\frac{\theta}{\sigma}[(1-q_1)H_1(t)+(1-q_2)H_2(t)]\right]=0.
\end{equation}

Next, dividing the cases: $q_1=q_2 \leq q^*$, $q_1<\min\{q_2, q^*\}$ and $q_2< q_1, q_2\leq q^*$ as in Theorem \ref{Example-1}, we derive the turnpike property (\ref{turnpike}) from (\ref{V-E-5}) and (\ref{V-E-7}).
 To derive (\ref{consumption}), noting $C(x,t)=-V'_2(u_x(x,t))$,  (\ref{V-C-C}), (\ref{V-E-5}) and $\lim_{t\to\infty} u_x(x,t)=\infty$, we have
$$ \lim_{t\to\infty} R(t)^{\frac{\hat q-1}{q_2-1}} C(x,t)
= \lim_{t\to\infty} \left(R(t)^{1\over q_2-1}u_x(x,t)\right)^{\hat q-1}= x^{\frac{ \hat q-1}{\min \{q_1,q_2\}-1}}.
$$
The last equality is derived by discussing three cases for $q_1$ and $q_2$, the same as that in deriving (\ref{turnpike}).
\qed

\bigskip\noindent
{\bf Proof of Theorem \ref{Main-1a}}.
The proof is similar to that of Theorem \ref{Main-1}. With strengthened assumptions, we can give better estimations to some inequalities in the proof of Theorem \ref{Main-1}. From (\ref{V-NC}) we conclude that assumption (\ref{V-C-3}) holds for all $y>0$ with a changed constant $L$. This implies that (\ref{V-E-2}) becomes
\begin{equation} \label{V-E-2b}
|H(t)-x|\leq LM(t),
\end{equation}
and (\ref{V-E-3}) becomes
\begin{equation} \label{V-E-3b}
H(t)\leq H:=(x+1+ L)^{\frac{q_1-1}{q_1-\bar{q}}}+\left[x +1+ \frac{LE^{\bar{q}-1 \over q_2-1}}{|\bar{\lambda}|} \right]^{\frac{q_2-1}{q_2-\bar{q}}}.
\end{equation}
 We have also the estimate (\ref{V-E-m}).
(\ref{V-E-6}) becomes
\begin{equation} \label{V-E-7b}
\left|A(x,t)-\frac{\theta}{\sigma}[(1-q_1)H_1(t)+(1-q_2)H_2(t)]\right|
\leq \frac{D\theta}{\sigma}L M(t),
\end{equation}
where $D= 1-\bar{q}+\frac{2Ee^{\lambda^+_2}+1}{a\sqrt{\pi}}$.
Note that $\tilde{q}< \min\{\hat{q}, q_2\}$, (\ref{V-C-C-C}) holds for all $y>0$. Setting $c:=\frac{\hat q -1}{\min\{q_1,q_2\}-1}$,
we have, from (\ref{e51})
\begin{equation}\label{consumption-1}
|R(t)^{\frac{\hat q-1}{q_2-1}} C(x,t)-x^c| \leq R(t)^{\frac{\hat q-1}{q_2-1}}|V'_2(u_x)+u_x^{\hat q-1}|+|R(t)^{\frac{\hat q-1}{q_2-1}} u_x^{\hat q-1}-x^c|.
\end{equation}
Noting that $R(t)^{\frac{\hat q-1}{q_2-1}}\leq (E(t))^{\frac{\hat q-1}{q_2-1}}$,
we have, from (\ref{V-C-C-C}) and (\ref{V-E-4})
\[
R(t)^{\frac{\hat q-1}{q_2-1}}|V'_2(u_x)+u_x^{\hat q-1}|\leq LR(t)^{\frac{\hat q-1}{q_2-1}} u_x^{\tilde{q}-1}\leq L H^{\frac{\tilde{q}-1}{\max\{q_1,q_2\}-1}}
(E(t))^{\frac{\tilde{q}-\hat q}{\min\{q_1,q_2\}-1}}.
\]
This gives the estimate for the first term in (\ref{consumption-1}).

For $q=q_1=q_2\leq q^*$, we deduce from (\ref{V-E-7b}) and (\ref{V-E-2b}),
\[
\left|A(x,t)-\frac{\theta}{\sigma}(1-q)x\right|
\leq \frac{\theta}{\sigma}L(1-q+D)M(t).
\]
 Let $G=G(x,c)=(1+c)H^{c-1}$ if $c>1$ and $G=G(x,c)=(1+c)x^{c-1}$ if $c\leq 1$.  It is clear that $x<H$ since $\frac{q_1-1}{q_1-\bar{q}}>1$ from (\ref{V-E-3b}).
Noting $R(t)u_x^{q-1}=H(t)$ and $x \leq \max\{H(t),x\} \leq H$, we deduce the estimate for the second term in (\ref{consumption-1}) by
(\ref{V-C-C-C}), (\ref{V-E-2b}) and
(\ref{e51}),
\[
|R(t)^c u_x^{\hat q-1}-x^c|
=|(R(t)u_x^{q-1})^c - x^c|\leq G(x,c)|H(t)-x|
\leq LG(x,c)M(t).
\]
We deduce convergence rate by (\ref{V-E-m}).

If $q_1<\min\{q_2, q^*\}$, we have, by (\ref{i-c}),
\[
H_2(t)\leq H^{\frac{q_2-1}{q_1-1}}te^{(\lambda^+_2-\lambda_1\frac{q_2-1}{q_1-1}) t},
\]
and by (\ref{V-E-2b}) and (\ref{V-E-7b}),
\[
\left|A(x,t)-\frac{\theta}{\sigma}(1-q_1)x\right|
\leq \frac{\theta}{\sigma}L(1-q_1+D)M(t)+(q_2-q_1)\frac{\theta}{\sigma}H_2(t).
\]
Note that $R(t)=e^{\frac{q_2-1}{q_1-1}\lambda_1 t}$ in this case, we have
$$
|R(t)^c u_x^{\hat q-1}-x^c| = |H_1(t)^c - x^c|
\leq G(x,c)|H_1(t) -x|
\leq G(x,c)(LM(t)+H_2(t)).
$$

If $q_2< q_1, q_2\leq q^*$, we have, by (\ref{c-i}), for $t\geq 1$
\[
H_1(t)\leq  (EH)^{\frac{q_1-1}{q_2-1}}e^{(\lambda_1-\lambda_2\frac{q_1-1}{q_2-1}) t},
\]
where $E:=\frac{\lambda_2}{1- e^{-\lambda_2}}$.
We deduce from (\ref{V-E-2b}) and (\ref{V-E-7b})
\[
\left|A(x,t)-\frac{\theta}{\sigma}(1-q_2)x\right|
\leq \frac{\theta}{\sigma}L(1-q_2+D)M(t)+(q_1-q_2)\frac{\theta}{\sigma}
H_1(t).
\]
Noting that $R(t)=\frac{e^{\lambda_2 t}-1}{\lambda_2}$, from (\ref{V-E-2b}),
\[
|R(t)^c u_x^{\hat q-1}-x^c| = |(H_2(t))^c -x^c|
\leq G(x,c)|H_2(t) -x|
\leq G(x,c)(LM(t)+H_1(t)).
\]
Note that $E(t)$ converges to 0 exponentially as $t\to\infty$ if $q_1<q^*$ or $q_2 < q^*$, and polynomially if $q_1\geq q^*$ and $q_2=q^*$. All other terms converge to 0 exponentially.
Combining all discussions above, we have proved the results.
\qed

\bigskip\noindent
{\bf Proof of Theorem \ref{Main-2}}.
In this case, $\lambda_1\leq 0$, $\lambda_2< 0$.
From (\ref{eqn10}) and (\ref{V-NC}), we conclude that there is constant $\bar{q}\in (\max\{q_1,q_2,0\}, 1)$, and
for any fixed $\epsilon>0$, there is $K_\epsilon>0$, such that
\begin{equation}\label{V-C-5-e}
\left|V'_i(y)+ y^{q_i-1}\right|\leq \epsilon y^{q_i-1} + K_\epsilon y^{\bar{q}-1}.
\end{equation}
Let $h(y,t)=\int_{0}^t I_2(y,\tau)d\tau$; then from (\ref{V-C-5-e}) ($i=2$)
\begin{equation}\label{h-1}
\left|h(y,t)-y^{q_2-1}\frac{e^{\lambda_2 t}-1}{\lambda_2}\right| \leq \epsilon y^{q_2-1}\frac{e^{\lambda_2 t}-1}{\lambda_2}+ K_\epsilon y^{\bar{q}-1}\frac{e^{\bar{\lambda} t}-1}{\bar{\lambda}}.
\end{equation}
Hence $h(y,t)$ is increasing and bounded above in $t$. We deduce that $\lim_{t\rightarrow \infty}h(y,t)=h(y)$, which implies that $h(y)$ is well defined and
\begin{equation}\label{h-2}
\left|h(y)-\frac{1}{|\lambda_2|}y^{q_2-1}\right|
\leq \epsilon\frac{1}{|\lambda_2|}y^{{q_2}-1}+  K_{\epsilon} \frac{1}{|\bar{\lambda}|}y^{\bar{q}-1}.
\end{equation}
Since $|V'_2(y)|$ is decreasing, we conclude that $h(y)$ is strictly decreasing, $h(\infty)=0$ and $h(0)=\infty$ by (\ref{h-2}). Hence equation (\ref{root}) admits a unique solution for any fixed $x>0$. We denote this unique solution by $Y=Y(x)$.

Since $h(y)-h(y,t)=\int_{t}^\infty I_2(y,\tau)d\tau$ and $yh'(y)-yh_y(y,t)=\int_{t}^\infty J_2(y,\tau)d\tau$, we have from (\ref{V-C-5-e}) ($i=2$)
\begin{equation}\label{h-3}
\left|h(y)-h(y,t)-\frac{1}{|\lambda_2|}y^{q_2-1}e^{\lambda_2 t}\right| \leq \epsilon y^{q_2-1}\frac{e^{\lambda_2 t}}{|\lambda_2|}+ K_\epsilon y^{\bar{q}-1}\frac{e^{\bar{\lambda} t}}{|\lambda|},
\end{equation}
and
\begin{equation}\label{h-4}
\left|yh'(y)-yh_y(y,t)-(1-q_2)y^{q_2-1}\frac{e^{\lambda_2 t}}{|\lambda_2|}\right|\leq (1-q_2+\frac{1}{a\sqrt{\pi}})\left[\epsilon y^{q_2-1}\frac{e^{\lambda_2 t}}{|\lambda_2|}+ K_\epsilon  y^{\bar{q}-1}\frac{e^{\bar{\lambda} t}}{|\lambda|}\right].
\end{equation}

By (\ref{V-C-5-e})(i=1) and (\ref{A}), we get
\[
\left|v_y(y,t)+y^{q_1-1}e^{\lambda_1 t}+h(y,t)\right|\leq \epsilon y^{q_1-1}e^{\lambda_1 t}+ K_\epsilon y^{\bar{q}-1}e^{\bar{\lambda} t}.
\]
Hence for fixed $x>0$
\begin{equation}\label{V-E-8}
\left|u_x(x,t)^{q_1-1}e^{\lambda_1 t}+h(u_x(x,t),t)-x\right|\leq \epsilon u_x(x,t)^{q_1-1}e^{\lambda_1 t} + K_\epsilon u_x(x,t)^{\bar{q}-1}e^{\bar{\lambda} t}.
\end{equation}
Note that $\lambda_1=0$ if $q_1=q^*$, $\lambda_1<0$ if $q_1>q^*$ and $\bar{\lambda}<0$. Taking $\epsilon=\frac{1}{2}$, we see from (\ref{V-E-8}) and (\ref{h-1}) that there exists a constant $\bar{Y}=\bar{Y}(x)$ such that
$u_x(x,t)\leq \bar{Y}$. Using (\ref{h-1}) and taking $\epsilon=\frac{1}{2}$, we deduce $\lim_{y\rightarrow 0} h(y,t)=0$
uniformly in $t$ when $t\geq 1$. Hence there exists a lower bound $\underline{Y}=\underline{Y}(x)$, $u_x(x,t)\geq \underline{Y}$.
Now assume, for any sequence ${t_k}$, $\lim_{k\rightarrow \infty}t_k=\infty$ and $\lim_{k\rightarrow \infty}u_x(x,t_k)=\hat{Y}$. Then (\ref{h-3}) yields $\lim_{k\rightarrow \infty}h(u_x(x,t_k),t_k)=h(\hat{Y})$. Letting $k\to\infty$ and then $\epsilon\to 0$ in (\ref{V-E-8}) with $t=t_k$, we conclude $\hat{Y}^{q_1-1}1_{\{q_1=q^*\}}+h(\hat{Y})=x$ and $\hat{Y}=Y$, by the uniqueness of solution for equation (\ref{root}). Therefore, $\lim_{t\rightarrow \infty}u_x(x,t)=Y$.
This yields (\ref{consumption-2}) directly.

By (\ref{V-C-5-e}), (\ref{PF-2}) and (\ref{A}), we get
\[
\left|yv_{yy}(y,t)-(1-q_1)y^{q_1-1}e^{\lambda_1 t}-yh_y(y,t)\right|
\leq  (1-q_1+\frac{1}{a\sqrt{\pi}})\left[\epsilon y^{q_1-1}e^{\lambda_1 t}
+ K_\epsilon y^{\bar{q}-1}e^{\bar{\lambda} t}\right],
\]
hence
\begin{eqnarray*}
&&\left|A(x,t)-\frac{\theta}{\sigma}[(1-q_1)(u_x(x,t))^{q_1-1}e^{\lambda_1 t}-u_x(x,t)h_y(u_x(x,t),t)]\right|
\\
&\leq& (1-q_1+\frac{1}{a\sqrt{\pi}})\left[\epsilon u_x(x,t)^{q_1-1}e^{\lambda_1 t}
+ K_\epsilon u_x(x,t)^{\bar{q}-1}e^{\bar{\lambda} t}\right].
\end{eqnarray*}
Noting $\lambda_2, \bar{\lambda} <0$, it yields (\ref{turnpike-2}) by (\ref{h-4}).

If  $V_2$ satisfies $V_2'(y)=-y^{q_2-1}$ with $q_2<0$, then a simple calculus shows that when $q_2>q^*$,
we have $ h(y)=-(1/\lambda_2)y^{q_2-1}$ and $yh'(y)=(q_2-1)h(y)$.
If $q_1>q^*$, then relations (\ref{turnpike-2})-(\ref{root})  give (\ref{eqn2.18}).
\qed

\bigskip\noindent
{\bf Proof of Theorem \ref{Main-3}}.
From (\ref{V-wC0}), for any fixed $0<\gamma<1-\max\{q_1,q_2\}$, there is $0<Y=Y_\gamma<1$ such that
\[
|R_i(y)-(1-q_i)|\leq \gamma,\ y\leq Y.
\]
Since $R_i(y)=-yV_i''(y)/V_i'(y)$, we have
$$
V_i'(y)=V_i'(Y)e^{\int_y^Y\frac{R_i(\eta)}{\eta}d\eta}
$$
for $y>0$. Hence, there exist positive constants $l=l_\gamma$ and $L=L_\gamma$, such that
\begin{equation}\label{V-E-11}
l y^{q_i+\gamma-1}\leq |V'_i(y)|\leq L y^{q_i-\gamma-1}, \,y\leq Y.
\end{equation}
We now give some estimates for $I_1(u_x,t)$ and $ \int_{0}^t I_2(y,\tau)d\tau|_{y=u_x}$. Suppose $q_1<\min\{q_2, q^*\}$, we choose $\gamma<\min\{\frac{q_2-q_1}{2}, q^*-q_1\}$. Then
\begin{eqnarray*}
I_1&\geq& \frac{e^{\beta t}}{2\sqrt{\pi }}\int_{-\infty}^{\frac{1}{a\sqrt{t}} \ln\frac{Y}{y}}
e^{-\frac{\eta^2}{4}-(\alpha-1) a\sqrt{t}\eta}|V_1'(ye^{a\sqrt{t}\eta})|d\eta\\
&\geq& \frac{l y^{q_1+\gamma-1}e^{\lambda(q_1+\gamma) t}}{2\sqrt{\pi }}\int_{-\infty}^{\frac{1}{a\sqrt{t}} \ln\frac{Y}{y}+2(\alpha-q_1-\gamma)a\sqrt{t}}
e^{-\frac{\xi^2}{4}}d\xi.
\end{eqnarray*}
 Suppose that $\ln\frac{Y}{y}+2(\alpha-q_1-\gamma)a^2 t\geq 0$; then we
get $x\geq I_1 \geq \frac{l}{2}u_x^{q_1+\gamma-1}e^{\lambda(q_1+\gamma) t}$. Otherwise, $\ln\frac{Y}{y}+2(\alpha-q_1-\gamma)a^2 t\leq 0$, we have $u_x \geq Y e^{2(\alpha-q_1-\gamma)a^2 t}$.
Noting that $\alpha-q\geq \alpha-q^*>0$ for $q\leq q^*$ and $\lambda(q)+2(q-1)(\alpha-q)a^2=-[a^2(q-1)^2+r]<0$,
we conclude that
\begin{equation}\label{V-wE-1}
u_x(x,t)^{q_1+\gamma-1}e^{\lambda(q_1+\gamma) t}\leq \max \{\frac{2x}{l},
Y^{q_1+\gamma-1} \}.
\end{equation}
Next assume $q_2<\min\{q_1, q^*\}$, we chose $\gamma<\min\{\frac{q_1-q_2}{2}, q^*-q_2\}$.
Similarly, we have
\[
\int_{0}^t I_2(y,\tau)d\tau\geq\int_{0}^t \frac{l y^{q_2+\gamma-1}e^{\lambda(q_2+\gamma)\tau}}{2\sqrt{\pi }}d\tau
\int_{-\infty}^{\frac{1}{a\sqrt{\tau}} \ln\frac{Y}{y}+2(\alpha-q_2-\gamma)a\sqrt{\tau}}
e^{-\frac{\xi^2}{4}}d\xi.
\]
 Let $k_0=\frac{\lambda(q_2+\gamma)}{2a^2(1 -q_2-\gamma )(\alpha-q_2-\gamma)}\in (0,1)$ and
$E_0=\frac{\lambda(q_2+\gamma)}{1-e^{(k_0-1)\lambda(q_2+\gamma)}}>0$. Assume $\ln\frac{Y}{y}+2(\alpha-q_2-\gamma)a^2 k_0 t\geq 0$, then $x\geq  \frac{l}{2E_0}u_x^{q_2+\gamma-1}e^{\lambda(q_2+\gamma) t}$ for $t\geq 1$. Otherwise, we conclude that $u_x \geq Y e^{2(\alpha-q_2-\gamma)a^2 k_0 t}$. We get, for $t\geq 1$,
\begin{equation}\label{V-wE-2}
u_x(x,t)^{q_2+\gamma-1}e^{\lambda(q_2+\gamma) t}\leq \max \{\frac{2\bar{E}x}{l},
Y^{q_2+\gamma-1}\}.
\end{equation}
Since $q_2+\gamma<q^*$, we have $\lambda(q_2+\gamma)>0$ and $e^{\lambda(q_2+\gamma)t}\to\infty$ as $t\to \infty$. Noting $q_2+\gamma-1<0$ and (\ref{V-wE-2}), we conclude that
$\lim_{t\rightarrow \infty}u_x(x,t)=\infty $. On the other hand, from (\ref{V-NC}), we obtain with a changed $L$
\[
|V'_i(y)|\leq  L(y^{q_i-\gamma-1}+y^{\bar{q}-1}), \ y>0.
\]
Hence we deduce upper bounds
\begin{equation}\label{V-wE-1a}
I_1\leq K[y^{q_1-\gamma-1}e^{\lambda(q_1-\gamma) t}+y^{\bar{q}-1}e^{\bar{\lambda} t}]
\end{equation}
and
\begin{equation}\label{V-wE-2a}
\int_{0}^t I_2(y,\tau)d\tau\leq K\left[y^{q_2-\gamma-1}\frac{e^{\lambda(q_2-\gamma) t}-1}{\lambda(q_2-\gamma)}+y^{\bar{q}-1}\frac{e^{\bar{\lambda} t}-1}{\bar{\lambda}}\right].
\end{equation}
Note that
\begin{equation} \label{V-wE-3}
yV''(y)=-R(y)V'(y)= (q-1)V'(y)+(1-q-R(y))V'(y).
\end{equation}
From (\ref{V-wC}) and (\ref{V-NC}), we conclude that
for any fixed $\epsilon>0$, there is $K=K_\epsilon>0$, such that
\begin{equation} \label{V-wE-3a}
|(1-q-R(y))V'(y)|\leq \epsilon |V'(y)| + K y^{\bar{q}-1}.
\end{equation}
Substituting (\ref{V-wE-3}) into (\ref{PF-2}), we get by (\ref{V-wE-3a})
\begin{eqnarray}
&&\left|y v_{yy}(y,t)-(1-q_1)I_1-(1-q_2)\int_{0}^t I_2(y,\tau)d\tau\right|\nonumber\\
&\leq& \epsilon\left(I_1+\int_{0}^t I_2(y,\tau)d\tau\right) + K_\epsilon y^{\bar{q}-1}\left(e^{\bar{\lambda} t}+\frac{e^{\bar{\lambda} t}-1}{\bar{\lambda}}\right). \label{V-wE-4}
\end{eqnarray}

Letting $y=u_x$,  $t\to\infty$ and then $\epsilon\to 0$ in (\ref{V-wE-4}), also noting $\bar{q}>q^*$, $\bar{\lambda}<0$ and $\lim_{t\rightarrow \infty}u_x(x,t)=\infty$, we obtain
\[
\lim_{t\rightarrow \infty} \left[A(x,t)-\frac{\sigma}{\theta}((1-q_1)I_1+(1-q_2)\int_{0}^t I_2(y,\tau)d\tau)\right]=0.
\]
If $q_1=q_2 < q^*$, we deduce (\ref{turnpike}) directly.
Assume $q_1<\min\{q_2, q^*\}$.  By (\ref{V-wE-1}) and (\ref{V-wE-2a}), we get
$\lim_{t\rightarrow \infty}\int_{0}^t I_2(u_x,\tau)d\tau=0$, then (\ref{turnpike}).
Assume $q_2<\min\{q_1, q^*\}$. By (\ref{V-wE-2}) and (\ref{V-wE-1a}), we get
$\lim_{t\rightarrow \infty}I_1=0$ and (\ref{turnpike}).
\qed

\bigskip\noindent
{\bf Proof of Theorem \ref{Main-4}}.
 From representation theorem for functions of regular variation (Bingham et al. (1987), Theorem 1.3.1 and equation (1.5.1)),
we conclude that (\ref{V-E-11}). Assume $q_1\neq q_2$, we chose $\gamma<\min\{\frac{|q_1-q_2|}{2}, 1-q_1, 1-q_2\}$ in (\ref{V-E-11}).
Hence, we get $\lim_{y\to 0} {V_2'(y)\over V'_1(y)} = 0$ if $q_1 < q_2$ and
$\lim_{y\to 0} {V_1'(y)\over V'_2(y)} = 0$ if $q_2 < q_1$.

We claim that, for any fixed $t_0>0$, $v_y(y,t_0)$ satisfies (\ref{V-wC0}) with index  $1-\min\{q_1,q_2\}$ and (\ref{V-wC}).
Suppose $q_1 <q_2 $. From Potter's bound (Bingham et al. (1987), Theorem 1.5.6.iii)(see alsoBack et al. (1999), equation (73)),     (\ref{PF-2}) and (\ref{V-NC}), for any fixed $t_0>0$, we get, by the dominated convergence theorem,
$\lim_{y\to 0} {v_y(y,t_0)\over V'_1(y)} = e^{\lambda_1 t_0}$
and $\lim_{y\to 0} {yv_{yy}(y,t_0)\over V'_1(y)} = (q_1-1)e^{\lambda_1 t_0}$.
We have
\[
\lim_{y\to 0}R(y,t_0):= {-yv_{yy}(y,t_0)\over v_y(y,t_0)} = 1-q_1.
\]
Hence $v_y(y,t_0)$ satisfies (\ref{V-wC0}) with index $1-q_1$. By (\ref{PF-2}), (\ref{V-NC}) and Potter's bound, we deduce (\ref{V-wC}).
Other cases can be prove similarly.
We have shown that $u_x(x,t_0)$ satisfies  (\ref{V-2}) with index $1-\max\{p_1, p_2\}$. Consider the following problem: find a solution $u=\bar{u}(x,t)$ satisfying the equation
\begin{equation} \label{HJBE-1b}
-{\partial u\over \partial t}-\frac{1}{2}\theta^2{u_x^2 \over u_{xx}}+rxu_x-\delta u+V_2(u_x)=0
\end{equation}
with  the initial condition $\bar{u}(x,0)=u(x,t_0)$ for $x\in R_+$, where $t$ represent a time horizon variable. Note that $u(x,t_0)$ is a utility function.
Then we get $\bar{u}(x,t)=u(x,t+t_0)$ and $\bar{A}(x,t)=A(x,t+t_0)$ for all $x\in R_+$ by the uniqueness of solution to the Cauchy problem of the Bellman equation. From Theorem \ref{Main-3} and the dual representation,
we deduce $\lim_{t\rightarrow \infty}\bar{A}(x,t)=\frac{\theta}{\sigma}(1-\min\{q_1,q_2\})x$, and the turnpike property (\ref{turnpike}) holds.
\qed

\bigskip
\noindent{\bf\Large References}

\begin{description}

\item Back, K., Dybvig, P.H., \&   Rogers, L.C.G. (1999). Portfolio turnpikes.  {\it Review of Financial Studies}, 12, 165-195.

\item Bian, B., Miao, S., \& Zheng, H. (2011). Smooth value functions for a class of nonsmooth utility maximization problems.  {\it SIAM Journal on Financial Mathematics}, 2, 727-747.

\item Bian, B., \& Zheng, H. (2015). Turnpike property and convergence rate for an investment model with general utility functions. {\it Journal of Economic Dynamics and Control}, 51, 28-49.

\item Bingham, N. H.,   Goldie, C. M., \&   Teugels, J. L. (1987). Regular Variation. Cambridge University Press, Cambridge, Mass.

\item Cox, J., \& Huang, C. (1992). A continuous time portfolio turnpike theorem.  {\it Journal of  Economic Dynamics Control}, 2, 491-507.

\item Guasoni, P., Kardaras, C., Robertson, S. \& Xing, H. (2014).  Abstract, classic, and explicit turnpikes. {\it Finance and Stochastics}, 18, 75-114.

\item Huang, C., \&  Zariphopoulou, T. (1999). Turnpike behaviour of long-term investments.  {\it Finance and Stochastics}, 3, 15-34.

\item Huberman, G., \& Ross, S. (1983). Portfolio turnpike theorems, risk aversion, and regularly varying functions. {\it Econometrica}, 51, 1345-1361.

\item Jin, X. (1998). Consumption and portfolio turnpike theorems in a continuous-time finance model. {\it Journal of Economic Dynamics and Control}, 22, 1001-1026.

\item Robertson, S. \&  Xing, H. (2017). Long term optimal investment in matrix valued factor models. {\it SIAM Journal on Financial Mathematics}, 8,  400-434. 

\end{description}

\end{document}